\documentclass[11pt,a4paper,oneside]{article}
\usepackage{hyperref} 
\usepackage{natbib}
\usepackage{graphicx}
\usepackage{color}
\usepackage{inputenc}
\usepackage{setspace}
\usepackage{geometry}
\usepackage{layout}
\usepackage{bbm}
\usepackage{amsmath}
\usepackage{amssymb}
\usepackage{makeidx}
\usepackage{tikz}
\usepackage{pgf}
\usepackage{xcolor}
\usepackage{titlepic}
\usepackage{rotating} 
\usepackage{float}
\usepackage[toc,page]{appendix}
\title{Interbank credit and the money manufacturing process. \\A systemic perspective on financial stability \\ --- \\ \large{Financial Regulation Research Lab (Labex ReFi) Working Paper, Paris}}
\author{Yuri Biondi\footnote{Senior Tenured Research Fellow, Cnrs – IRISSO (UMR7170, University Paris Dauphine PSL), Paris, France. Research Director, Labex ReFi, Paris, France. (email: yuri.biondi@gmail.com)} \ and Feng Zhou\footnote{Department of Computer Science and Kellogg College, University of Oxford, Oxford, UK. Post-Doctoral Researcher, Labex ReFi, Paris, France. (email: feng.zhou@kellogg.ox.ac.uk)}}
\begin{document}
\maketitle
\begin{abstract}
Interbank lending and borrowing occur when financial institutions seek to settle and refinance their mutual positions over time and circumstances. This interactive process involves money creation at the aggregate level. Coordination mismatch on interbank credit may trigger systemic crises. This happened when, since summer 2007, interbank credit coordination did not longer work smoothly across financial institutions, eventually requiring exceptional monetary policies by central banks, and guarantee and bailout interventions by governments. Our article develops an interacting heterogeneous agents-based model of interbank credit coordination under minimal institutions. First, we explore the link between interbank credit coordination and the money generation process. Contrary to received understanding, interbank credit has the capacity to make the monetary system unbound. Second, we develop simulation analysis on imperfect interbank credit coordination, studying impact of interbank dynamics on financial stability and resilience at individual and aggregate levels. Systemically destabilizing forces prove to be related to the working of the banking system over time, especially interbank coordination conditions and circumstances.\\
\\
\textbf{Highlights}
\begin{itemize}
	\item An interacting heterogeneous agents-based model under minimal institutions is developed to study bank credit, money creation and interbank credit
	\item Interbank credit makes the money generation process unbound in some circumstances
	\item Financial stability depends on both bank and inter-bank credit conditions and circumstances
\end{itemize}
\\ 
\textbf{Acknowledgements:} Yuri Biondi is tenured senior research fellow of the National Center for Scientific Research of France (Cnrs - IRISSO), and research director at the Financial Regulation Research Lab (Labex ReFi), Paris, France. Feng Zhou is post-doctoral researcher at the Financial Regulation Research Lab (Labex ReFi), Paris, France.
\end{abstract}

\onehalfspacing
%\singlespacing
%%%%%%%%%%%%%%%%%%%%%%%%%%%%%%%%%%%%%%%%%%%%%%%%%%%%%%%%%%%
%\frontmatter 
%\tableofcontents
%\listoftables
%\listoffigures
%\citet{Anwar2012}
%\citep{Anwar2012}

%%%%%%%%%%%%%%%%%%%%%%%%%%%%%%%%%%%%%%%%%%%%%%%%%%%%%%%%%%%
\section{Introduction}
During recent decades before the global financial crisis, mainstream economic theory had quite neglected the role of money and credit in economy and society \citep{jakab2015banks, werner2014can, werner2014banks}. In particular, macroeconomic theory was drawing upon the real business cycle approach, developing Dynamic Stochastic General Equilibrium (DSGE) models where money and credit do not play a significant role \citep{Blanchard2009, stiglitz2011heterogeneous, romer2016trouble, DelliGatti20101627, Blanchard2010Rethinking}. In this context, banking may be understood as a mechanic process that merely dispatches central bank base money to non-financial, real-economic agents who need borrowing. Moreover, banking theory was developing principal-agent approaches that introduce contractual arrangements and incentives in the bank entity structure while relegating bank money generation function outside banking theoretical core \citep{Calorimis1991, diamond2001liquidity, gorton1990financial}. At the same time, financial economics was developing a market-based finance theory where banks are considered as portfolio managers quite analoguous to other investment funds and submitted to financial markets discipline \citep{tobin1963commercial, Black1970, fama1980banking, hall1983optimal}. From these perspectives, banking come to be understood as pure intermediation of available  funds between  savers who wish to invest those funds, and borrowers  who need borrowing  them.\\
\\
Evidence from the global financial crisis of 2007-08 has been renewing theoretical interest in the role of banking, money and credit in economic and financial dynamics. While the crisis was accidentally triggered by rising actual and expected defaults in some bank asset categories, the origins of the global financial meltdown were found in systemic fragilities related to bank excess behaviours and the overarching dynamics of interbank credit, leading to central  bank interventions and government bailouts to protect financial stability and assure financial resilience \citep{Bank2008}. Interbank lending and borrowing occur when financial institutions seek to settle and refinance their  mutual positions over time and circumstances. This interactive process involves money generation at the aggregate level. Coordination mismatch on interbank credit may trigger systemic crises. This happened when, since summer 2007, interbank credit coordination did not longer work smoothly across financial institutions around the world, eventually requiring exceptional monetary policies through central bank coordination.\\
\\
Recent studies have been investigating the role of money, credit and interbank credit networks in the working of the financial system, including when financial crises occur. \citet{Martinez20102358} and \citet{Caccioli201550} develop simulation models that relate interbank dynamics with systemic risk. \citet{Matsuoka20121673} provides  a monetary model for understanding the  role of a lender  of last  resort (i.e.:  central  banks)  in an economy with  an imperfect  interbank market.     \citet{Anand20121088} address  the heavy reliance on short-term wholesale funding markets  in a vastly  and increasingly  connected financial system  during  the  global financial crisis of 2007/2008,  leading to a dramatic increase in rollover risk at  the system  level.   Further studies focus on the  interbank credit network  structure and  the financial  linkages  between  banks,  often  applying agent-based simulations \citep{Krause2012583, Teteryatnikova2014186, Bech201544, Capponi2015152}.\\
\\
Many studies apply agent-based modelling (ABM) to analyze the interbank credit dynamics. \citet{Huang20101105} apply an ABM to examine various types of financial crises. \citet{Galbiati2011859} propose an ABM of interbank payment systems under real-time  gross settlement modality that seems to be vulnerable to liquidity risk. To their knowledge, it is the first paper that explores liquidity management in such a system using an agent-based approach. Other ABM are applied to explain: the formation, evolution and stability of interbank market \citep{Iori2006525, Xu2016131, Ladley20131384}, the emergence of network structure (e.g.: core-periphery) \citep{Lux2015A11}, the linkage between the interbank market and the real economy \citep{Gabbi2015117}, the repercussions of inter-bank connectivity on banks' performances, bankruptcy waves and business cycle fluctuations \citep{Grilli2015}, the role of trading memory or trust in the interbank relationships \citep{Iori201598, Temizsoy2015118, bulbul2013determinants}, the behavior of bank leverage \citep{Fischer201495, Aymanns2015155}, the  role of stabilizing institutional arrangements based  on socioeconomic  roles and  leaderships \citep{Gilles2015375}, the bank default and resolution \citep{Klimek2015144}. More generally, \citet{Bargigli20141} introduce some basic concepts and instruments in a wide range of economic network models.\\
\\
Pointing to the link between the banking system and real business economy,  \citet{DelliGatti20101627} model a more general  credit  network  consisting  of households,  firms and  banks, in view to study the occurrence and likelihood of bankruptcy avalanches. \citet{Battiston20121121}, \citet{Riccetti20131626}, \citet{Bargigli2014109}, and \citet{Catullo201578}  extend  this  work.  \citet{Lux201636} and \citet{Anand2013219} develop  stochastic  models of the topology  of bank-firm  credit  networks.  \citet{He20161} construct an  endogenous credit  network  model that describes the formation  of firm-firm, firm-bank  and bank-bank credit  relationships.   Moreover, \citet{Gatti2015}  and \citet{Beck20141} apply  ABM to explore the effects of monetary policy. Their  simulations  show its clear non-neutrality and transmission mechanism  in the credit  channel.  \citet{Valencia201420} discuss a similar issue.    \citet{Gabbi2015117} focus on the  linkage  between  the  inter- bank market  and the real economy with a stylized central  bank acting as a lender of last resort. Our article contributes and expands on these analyses by paying specific attention to the way banks perform credit creation along with inter-bank credit, clearing, and payment settlement.
\\
\\
Concerning modeling the money creation process,  \citet{Xiong2017425} develop a multi-agent model describing the main mechanisms of money creation and money circulation  in a credit economy. Similarly, \citet{chen2014money} examine the money creation process in a random exchange model with debt among agents. Our article expands on these analyses by combining payment and credit systems, while introducing endogenous money generation through bank credit and interbank credit.\\
\\
Our article contributes to this recent literature by taking a systemic perspective, aiming to overcome received limited understanding of the systemic links between money, credit and banking. Some facts feature modern banking: monetary financial institutions (banks) issue claims which function as money; they facilitate payments across agents in the economy over time and space; they increase the money base through credit creation; they hold fractional reserves and lend to each other \citep{Blair2013, ricks2016money, jakab2015banks}. Our systemic perspective points to these featuring dimensions of ongoing bank activity within each bank entity and across them. Each bank entity keeps currency money in bank deposits on behalf of other agents. But the bank entity activity is further characterized by its capacity or privilege to use these deposits, although the latter remain available for payment and redemption at will and at par. Moreover, the bank can create a deposit by granting a loan to, or buy a security from a borrower. This bank capacity or privilege involves money multiplication that enables bank credit creation process over space and time. In this way, all the banks become interdependent on the flow of payments that are performed across them, generating a `banking system' \citep{Yuri2017Banking}.\\
\\
This banking dynamic system requires coordination within each bank and across them. Within each bank entity, two featuring processes are at work: (i) An economic process that creates bank money through credit, in view to generate income to the bank entity; and (ii) a financial process that rebalances cash inflows and outflows when they become due through space and time. Since each bank is structurally unbalanced due to money multiplication, interbank coordination is required to maintain the banking system in operation over time and circumstances. Interbank clearing, settlement and credit arrangements feature this interbank coordination which may occur under various institutional arrangements, such as central banking, central clearing parties, and money markets. From this perspective, our conceptual framework embeds heterogeneous agents into two collective dynamics: inter-agent interaction, and interaction between collective structures and individual agents. These structures are consistent with the notion of `minimal institution' introduced by \citet{shubik2011note} and \citet{shubik2016guidance}.\\
\\
The rest of the article is organised as follows. Drawing upon this conceptual frame of reference, we develop an interacting heterogeneous agents-based model of interbank credit coordination under a set of minimal institutions (Section 2). First, we explore the theoretical link between interbank credit coordination and the money generation process (Section 3). Second, we develop simulation comparative analysis on modes of interbank credit coordination, studying impact of interbank dynamics on financial stability and resilience at individual and aggregate levels (Section 4). Systemically destabilising forces prove to be related to the working of the banking system over time, especially interbank coordination conditions and circumstances. A summary of main findings and implications concludes (Section 5).

\section{Agents-based dynamic modelling of the banking system}
According to \citet{Xiong2017425}, the textbook story of money creation tells that the quantity of loans commercial banks could possibly grant is constrained by the quantity of central bank base money (central bank reserves, comprising currency and drawing facilities) and the required reserve ratio. This story is consistent with fractional reserve lending of loanable funds, where, period through period, each lending bank department is restricted to lend out only a fraction of available cash in hand. Under fractional reserve requirement, money is therefore created through a gradual, ad infinitum mechanical process of `lending out' and `depositing in' of cash in hand. This modelling strategy has been often criticised to neglect space and time of money creation process, implying a reductionist view on its idiosyncratic, interactive, collective and dynamic dimensions. In particular, this view makes banking system coordination virtually irrelevant.\\
\\
Our model extends and upgrades on this received understanding by drawing upon the functional equivalence between currency and deposit. From this functional perspective, money can be created both by issuing cash and cash-equivalent securities (that is, financial entitlements that circulate and function like cash does), and by creating drawing facilities (bank deposits) that promise to be redeemable in cash and cash-equivalents. In this context, banking, money and credit are fundamentally linked one to another. When a bank lends to a borrowing client, bank balance sheet simultaneously expands with the loan (an asset to the bank) and the deposit (a liability to the bank), both relating to that customer, while the bank promises to make the customer deposit redeemable in cash and cash equivalents (especially other banks' deposits). When the borrowing client repays its loan to the bank, the loan capital instalment reduces the customer's exposure (an asset reduction to the bank), while the bank does either acquire a different kind of asset (if settlement is performed by cash transfer), or reduce its deposit liability (if settlement is performed through customer's deposit).\\
\\
Our story replaces then fractional reserve lending with a money multiplication process. The bank treasury department seeks to rebalance movements in cash and cash equivalents that become due for settlement-related and credit-related payments, while the bank lending department levers upon cash and cash-equivalent holdings to create new loans. To disentangle the systemic connection between payments, bank credit and inter-bank credit, we develop an agent-based model comprising three functional steps that each bank passes through in managing its relation with customers and with the other banks. These dynamic steps correspond to ongoing dimensions of the banking system: (i) payments system (Section 2.2), (ii) bank credit creation and destruction with customers (Section 2.3), and (ii) interbank settlement and credit system (Section 2.4).

\subsection{Setting the scene (INITIALISATION)}
Our miniature financial system comprises the following agent types:
\begin{description}
	\item[$\diamond$] A whole of customers $c_j, j = 1...C$, which order payments over time periods t, while asking for bank loans and repay them over time periods $h>=t$. For sake of simplicity, we may assume that $h=t$ without generality loss.
	\item[$\diamond$] A system of monetary financial institutions (banks) $b_i, i = 1...B$, which perform payments on behalf of other agents, and grant credits to customers and to each other. 
	\item[$\diamond$] One Central Bank, which issues an exogenous quantity of currency $A1$ and acts as guarantor of last resort if one bank remains exposed to other banks after interbank settlement and credit.
\end{description}
Each bank balance sheet contains five kinds of assets and related liabilities (Table~\ref{table:BankBalanceSheet}).
%\subsection{Bank balance sheet}
\begin{table}[htbp]
\caption{Bank balance sheet items} % title of Table
\centering % used for centering table
\begin{tabular}{|l | l|} % centered columns (3 columns)
\hline
Asset       & Liability      \\
\hline %inserts double horizontal lines
A1: Currency Reserves & L1: Deposit (Base Money) \\
A2: Retail Loan	& L2: Deposit (Retail Loan)      \\
A3: Interbank Lending & L3: Interbank Borrowing      \\
A4: Equity Reserve & L4: Equity Provision     \\
A5: Central Bank Assistance	& L5: Central Bank Guarantee      \\
\hline %inserts single line
\end{tabular}
\label{table:BankBalanceSheet} % is used to refer this table in the text
\end{table}
%\onehalfspacing
\\
Bank asset side comprises:
\begin{itemize}
\singlespacing
	\item Currency Reserves (base money) held on behalf of customers, $A1_{i,t}$
	\item Loan granted to, or securities underwritten from customers, $A2_{i,t}$
	\item Loan granted to, or securities underwritten from other bank, $A3_{i,t}$
	\item Equity Reserve, $A4_{i,t}$
	\item Central Bank Assistance, $A5_{i,t}$
\end{itemize}
Bank liability side comprises:
\begin{itemize}
\singlespacing
	\item Currency Deposit by customers, $L1_{i,t}$
	\item Loan Deposit by customers, $L2_{i,t}$
	\item Interbank Loan Deposit by other banks, $L3_{i,t}$
	\item Initial Equity Provision plus cumulated accrued profits and losses, $L4_{i,t}$
	\item Central Bank contingent Guarantee, $L5_{i,t}$
\end{itemize}
\onehalfspacing
%\singlespacing
The initial allocation of base money between customers and banks is based on the following steps:
\begin{description}
	\item[Allocate every customer j to any one bank i randomly ($l1_{i,j}$ is j's cash deposit in i):] 
  \begin{equation}
	\{c_j\} \longrightarrow \{M_{c_j\rightarrow b_i}\} \longrightarrow \{b_i\}=l1_{i,j,t}
	\end{equation}
	\item[Allocate base money to all customers equally:] 
	\begin{equation}
	l1_{i,j,t=0}=A1_0/C
	\end{equation}
	\item[Deposit all customers' money to their own banks:]
	\begin{equation}
  A1_{i,t=0} \ = \ L1_{i,t=0} \ = \sum_{j\in m_{j,i}=1}{l1_{i,j,t=0}}, \ \ \ \forall m_{j,i}\in M_{c_j\rightarrow b_i}
  \end{equation}
	\item[Allocate total Equity Provision to banks equally:] 
	\begin{equation}
	A4_{i,t=0}=L4_{i,t=0}=A4_0/B
	\end{equation}
	\item[Other balance sheet items]
	\begin{equation}
	\begin{array}{c}
  A2_{i,t=0}=L2_{i,t=0}=0\\
  A3_{i,t=0}=L3_{i,t=0}=0\\
	A5_{i,t=0}=L5_{i,t=0}=0
  \end{array} 
	\end{equation}	
	\item[Bank balance sheet identity:] 
  \begin{equation}\label{eq:accountingequation1}
	\begin{array}{c}
  A1_{i,t=0}+A2_{i,t=0}+A3_{i,t=0}=L1_{i,t=0}+L2_{i,t=0}+L3_{i,t=0}\\
  A4_{i,t=0}=L4_{i,t=0}\\
	A5_{i,t=0}=L5_{i,t=0}
  \end{array} 
	\end{equation}
\end{description}
Based on the accounting equation~\ref{eq:accountingequation1} (balance sheet identity) of an individual bank $i$ at initial time period $t=0$ (a time period $t$ can be daily, weekly or monthly interval), we have the following balance at all time periods: 
\begin{equation} 
\begin{array}{c}
A1_{i,t}+A2_{i,t}+A3_{i,t}=L1_{i,t}+L2_{i,t}+L3_{i,t}\\
A4_{i,t}=L4_{i,t}\\
A5_{i,t}=L5_{i,t}
\end{array} 
\end{equation}
This miniature economy has a total base money $A1=\sum^{B}_iA1_{i,t=0}=\sum^{B}_i\sum^{C}_jl1_{i,j,t=0}$ that is distributed equally to $C$ customers, and a total bank capital $\sum^{B}_iL4_{i,t=0}$ that is distributed equally to $B$ banks. Customers deposit all of their money to individual banks, based on a fixed transition/selection matrix $M_{C\rightarrow B}=(m_{c_j\rightarrow b_i})_{C\times B}$ (where $m_{c_j\rightarrow b_i}=1$ if $b_i$ is the bank of $c_j$, else $m_{c_j\rightarrow b_i}=0$). At time $t=0$, the initial capital $L4_{i,t=0}$ of bank $i$ is provided, which also adds to its initial equity reserve $A4_{i,t=0}$ (the size of banks can be modeled equally or follow some distributions in line with real industry data). The initial total deposit from customers $c_{j, j\in \{C\}}$ to their bank $b_{i,i\in \{B\}}$ is $L1_{i,t=0}=\sum_jl1_{i,j}$ where deposited currency is issued by the central monetary authority (assuming all customers' money is stored in banks, no cash in hand outside the banking system), adding to the initial base money cash reserve $A1_{i,t=0}$ of bank $i$ (the initial money of customers can be modeled equally or follow some power law distributions in line with real data). There is no initial loan to customers $A2_{i,t=0}=0$ or $L2_{i,t=0}=0$, no interbank lending $A3_{i,t=0}=0$ or borrowing $L3_{i,t=0}=0$, no central bank assistance $A5_{i,t=0}=0$ or guarantee $L5_{i,t=0}=0$. Therefore, the initial individual bank balance sheet identity is:
\begin{equation} \label{eq:accountingequation2}
\begin{array}{l}
A1_{i,t=0}=L1_{i,t=0}\\
A4_{i,t=0}=L4_{i,t=0}
\end{array} 
\end{equation}
where $A1_{i,t=0}=L1_{i,t=0}=\sum_jl1_{i,j,t=0}$ and $l1_{i,j,t}$ is a customer j's individual cash deposit in his/her own bank i at time t.\\
\\
Since our analysis focuses on financial stability implications of interbank coordination, we assume that equity (A4/L4) and central bank (A5/L5) provisions are non-cash based, that is, they do not modify reserve base holdings and obligations over time and circumstances. A further extension may consider an additional set of minimal institutions to design and implement equity payments and central banking monetary interventions.\\
\\
Concerning bank equity capitalisation, for same of simplicity and in line with \citet{DelliGatti20101627}, we fix the number of banks and customers (i.e. exogenous) and we allocate equity capital equally across banks. Simulations start from identical initial conditions, but agents become rapidly heterogeneous after interactions. In some circumstances, banks may experience losses and then exhaust their initial equity provision. For sake of simplicity, we let survive banks with negative equity, using total equity loss as an indicator of financial distress for the single bank and the financial system as a whole. On the contrary, \citet{DelliGatti20101627} replace defaulted banks with new banks having small capital relative to the size of other solvent banks.\\
\\
The model further disentangles ongoing banking activity through as sequence of three dynamic modules: (1) \textit{Payments flow} (money circulation among customers, where banks facilitate payments over space and time); (2) \textsl{Lending to real economy} (banks are loan creators through long-term retail loans); (3) \textsl{Interbank coordination} (through both interbank credit and central bank assistance as a residual balance item).

\subsection{Model Module 1: Payments system}
In our miniature economy, customers $c_j$ may use currency $A1$ to perform payments across them. When this occurs, the bank of the customer $c_j$ which orders to pay does transfer customer currency reserve $A1_{j,t}$ while reducing customer currency deposit $L1_{j,t}$. Conversely, the bank of the customer which receives the payment does increase the customer currency deposit $L1_{j,t}$ and customer currency reserve $A1_{j,t}$. In this way, banks act as passive fiduciary depositors of customers’ holdings of currency.\\
\\
Moreover, due to functional equivalence between cash and bank deposit, customer $c_j$ may use their bank deposit $L2_{j,t}$ to perform payments. Customer bank deposits are created when their bank grants a loan $A2_{j,h}$ to them. When this bank deposit wire transfer occurs, the bank of the customer $x$ which orders to pay does reduce the customer bank deposit $L1_{x,t}$ by requested payment amount. Conversely, the bank of the customer $y$ which receives the payment does increase the customer bank deposit $L1_{y,t}$. However, by assumption, banks do not transfer currency reserves against bank deposit transfers. Instead, banks agree to grant interbank lending and borrowing of same amount. The payer's bank then borrows from the payee's bank, adding to its interbank liability $L3_{x,t}$, while the payee's bank lends to its paying counterparty, adding to its interbank loan portfolio $A3_{y,t}$. This mechanism reproduces the settlement mechanism that is generally in place for banks through central bank facilities \citep{rule2015understanding}.\\
\\
Initial exogenous quantity of central bank currency $A1_0$ was equally distributed across customers and it now moves stochastically across them through time periods $t$. All cash holdings $A1$ are held by financial institutions. Payments across customers settled in currency are generated through a random markov transition matrix across customer bank accounts. \\
\\
Customers perform two kinds of payments in this module: (1) cash payments through currency transfers from the banks of paying customers to the banks of receiving customers (i.e. updating both $A1_{i,t}$ and $L1_{i,t}$ in the banks' balance sheet); (2) bank wire transfers that pay through bank deposits $L2_{i,t}$ rather than transferring cash.

\subsubsection{Customers' cash payment (A1/L1 update)}
When modeling the first kind of payment (i.e. cash), a stochastic Markov transition matrix $\widetilde{M}_{c_x\rightarrow c_y}$ is used to describe the transitions of cash payment among $C$ customers. Each of its entries $\tilde{m}_{x\rightarrow y}$ represents the probability of a customer $c_x, x\in C$ to pay another customer $c_y, y\in C$, therefore $0\leq \tilde{m}_{x\rightarrow y}\leq 1$ and $\sum_y{\tilde{m}_{x\rightarrow y}}=1$. (This means that the sum of each row in $\widetilde{M}_{c_x\rightarrow c_y}$ is the total pay-out of a customer's existing cash deposit, and the sum of each column is the total pay-in from other customers that adds to a customer's total new deposit.) A parameter $\xi_1\in [0,1]$ controls the scale of customers' cash payment. This implementation implicitly assumes that customers are constrained by their cash deposit. Also, their net cash flow (a balance between cash payable and receivable) is both small and balanced over time.\\
\\
Hence, among the customers:
	\begin{equation}
	\{c_x\} \stackrel{\xi_1}{\longrightarrow} \{\widetilde{M}_{c_x\rightarrow c_y}\} \longrightarrow \{c_y\} \Rightarrow \Delta l1_{x\rightarrow y,t}
	\end{equation}
In parallel, the bank of customer $c_x$ transfers an amount $\Delta l1_{x\rightarrow y,t}$ of A1 to the bank of customer $c_y$, and reduces the cash deposit of $c_x$ accordingly. In aggregate, every bank performs netting between total cash-in and cash-out, updating both sides of their balance sheet as follows:  
	\begin{equation}
  \Delta A1_{i,t} = \Delta L1_{i,t} = \sum_{y\in \{\tilde{m}_{y\rightarrow x}>0\}}\Delta l1_{y\rightarrow x,t} \ \ - \sum_{x\in \{\tilde{m}_{x\rightarrow y}>0\}}\Delta l1_{x\rightarrow y,t}
	\end{equation}	
	where $c_x$ is the customer of bank $i$ and $c_y$ may be any customer of other banks. 
	
\subsubsection{Customers' bank credit wire transfer (L2 and A3/L3 update)}	
When modeling the second kind of payment (i.e. bank loan deposit wire transfer), a similar stochastic Markov transition matrix is used to describe the wire transfer payment transitions. In this case, customers use their loan deposits $l2_{i,j,t}$ to perform their payments.\\
\\
Therefore, among the customers:
	\begin{equation}
	\{c_x\} \longrightarrow \{\widetilde{M}_{c_x\rightarrow c_y}\} \longrightarrow \{c_y\} \Rightarrow \Delta l2_{x\rightarrow y,t}
	\end{equation}
However, there are two differences when comparing to the cash payment movement: (1) Although the payments are exchanged among customers, the actual net of credit transactions are among banks. Therefore, a new transition matrix $\widetilde{M}_{b_u\rightarrow b_v}$ is used to model the interbank credit net transition (i.e.: the net of total customer deposits in and out). (2) Since the customers can borrow from their banks to pay other customers via bank credit, the customers are constrained by the loan $l2_{i,j,t}$ granted by their own banks. A parameter $\xi_2\in [0,1]$ controls the scale of this transaction.
	\begin{equation}
	\{b_u\} \stackrel{\xi_2}{\longrightarrow} \{\widetilde{M}_{b_u\rightarrow b_v}\} \longrightarrow \{b_v\} \Rightarrow \Delta L2_{u\rightarrow v,t}
	\end{equation}	
where $\Delta L2_{u\rightarrow v,t}\leq L2_{u,t}$ for all banks. Each couple of banks perform clearing and settlement of these wire payments in two steps. First, the two banks net mutual opposite positions against each other. Second, they agree to transform the net residual difference in interbank lending and borrowing. \\
\\
Therefore, the netting transaction between any couple of two banks $(u,v)$ is as follows:
	\begin{equation}
  \Delta L2_{u,v,t} = \Delta L2_{u\rightarrow v,t} - \Delta L2_{v\rightarrow u,t}, \ \ \ \forall (u,v),t
	\end{equation}
where each bank coupled $(u,v)$ deposit movement from bank to bank is the sum of customers in one bank paying to customers in the other bank. For instance, for payments from bank $u$ to bank $v$: $\Delta L2_{u\rightarrow v,t}=\sum_{c_x\in b_u, c_y\in b_v}{\Delta l2_{x\rightarrow y,t}}$. In aggregate, every bank sums up all of these net wire payments among other banks and update its $L2_{i,t}$:
	\begin{equation}
  \Delta L2_{u,t} = \sum_{v\in \{\tilde{m}_{u\rightarrow v}>0\}}{\Big(\Delta L2_{u\rightarrow v,t} - \Delta L2_{v\rightarrow u,t}\Big)}
	\end{equation}
if $\Delta L2_{u,t}>0$, then it increases the bank u's L2, else if $\Delta L2_{u,t}<0$, then it decreases the bank u's L2. In parallel, banks update their interbank credit (A3 and L3) among them: 
\begin{equation}
\left\{
\begin{array}{ll}
\Delta A3_{u,t} = \Delta L2_{u,t},  & \ \ \ \mbox{if } \Delta L2_{u,t}>0 \\
\Delta L3_{u,t} = -\Delta L2_{u,t}, & \ \ \ \mbox{if } \Delta L2_{u,t}<0.
\end{array} \right.
\end{equation}
These new amounts of interbank credit are recorded as a pair $(u,v)$ between every two banks, and add to the other A3 and L3 that will be created in the interbank network and repaid together later on.		

\subsection{Model Module 2: Bank credit creation and destruction }
While dealing with the payments system on behalf of its customers, each and every bank experiences that an ongoing core of its reserve holdings $A*_{i,t}\in (A1_{i,t},A2_{i,t},A3_{i,t})$ remains relatively stable over time. Under the functional equivalence between currency and deposit, every bank acquires the capacity or the privilege to use these holdings, while they remain available to the agents that have borrowed from the bank, being them customers ($L1_{j,t}, L2_{j,t}$), or other banks ($L3_{j,t}$). This means that the bank holds an ongoing safety net on which its lending activity may lever upon.\\
\\
For sake of simplicity and without loss of generality, we assume that each bank lending department does plan to lend to its customers only. Before seeking to grant new loans, each bank receives loan repayments from customers which borrowed in the past. 

\subsubsection{Customers' loan repayments (A2/L2 update)} 
In line with bank loan repayment mechanism introduced by \citet{Xiong2017425}, customers loan repayment follows the following procedure. Generating a stochastic repayment ratio $\widetilde{\Psi}_{i,t}$ for each bank based on a Triangular Distribution (depending on the lower $\psi_L$, peak $\psi_P$ and upper $\psi_U$ parameters):
	\begin{equation}
	\widetilde{\Psi}_{i,t}(\psi_L,\psi_P,\psi_U) \sim Triangular(\psi_L,\psi_P,\psi_U)
	\end{equation}
	The loan is repaid based on the following repayment formula:
	\begin{equation}\label{eq:LoanRepayment}
	Repayment \Delta L2_{i,t} = \widetilde{\Psi}_{i,t}\cdot L2_{i,t}
	\end{equation}
This statistical distribution implies an average loan outstanding time of $3/(\psi_L+\psi_P+\psi_U)$ and a modal loan outstanding time of $\psi_P$. Since banks only lend to own customers, this mechanism implies simple reverse lending as both $A2_{i,t}$ and $L2_{i,t}$ are reduced by $Repayment \Delta L2_{i,t}$ in the bank balance sheet, where $L2_{i,t}$ is the existing loan portfolio of the bank. This mechanism treats the real economy as a whole, without tracking individual customers and then individual loans granted to them. In principle, each loan has its terms and conditions, including its own duration and repayment profile. This involves that, at each period of time, each bank is confronted with a portfolio of loans of different durations and repayment profiles. To capture this feature without tracking each loan, we move from a universal repayment ratio for every bank as in \citet{Xiong2017425} to a stochastic one for each bank. Its statistical distribution denotes the business cycle conditions in the real economy. \\
\\
The model is calibrated to avoid customers’ default on their loans, implying that $L2_{i,t}\geq \Delta L2_{i,t}, \forall i,t $ (as in Equation~\ref{eq:LoanRepayment} above). A further extension may consider an additional set of minimal institutions to feature customer default resolution arrangements \citep{goodhart2016macro}.

\subsubsection{Customer loan creation}
After being repaid for previous loans, banks seek to grant new loans to their customers. The model denotes two distinctive lending strategies, namely money multiplication and fractional reserve lending. \\
\\
\underline{Lending Behavior 1:} Bank money multiplication lending mechanism\\
\\
Based on the target reserve ratio $\gamma^{TR}_{i,t}$ and the existing ``total reserve holding'' $\sum{A*_{i,t}}$, the maximum of total credit that each bank can grant is:
	\begin{equation}
	\gamma^{TR}_{i,t} = \frac{\sum{A*_{i,t}}}{\mbox{Maximum Lending}} \ \ \Rightarrow  \ \ \mbox{Maximum Lending} = \frac{\sum{A*_{i,t}}}{\gamma^{TR}_{i,t}}
	\end{equation}
The existing total deposit is $\{L1+L2+L3\}_{i,t}$, then each bank wishes to lend a multiple of ``total reserve base'' to keep new and existing loans below the maximum level, so
	\begin{equation}
	\Delta A2_{i,t+1} \leq \frac{\sum{A*_{i,t}}}{\gamma^{TR}_{i,t}} - \{L1+L2+L3\}_{i,t} 
	\end{equation}
Therefore, the general formula of target lending under money multiplication is: 
	\begin{equation}\label{eq:LendingB2E0}
	Potential \Delta A2_{i,t+1} = Max\Bigg\{ \ 0 \ ,\ \ \frac{\sum{A*_{i,t}}}{\gamma^{TR}_{i,t}} - (L1_{i,t}+L2_{i,t}+L3_{i,t}) \ \Bigg\}
	\end{equation}		
This formula implies that banks wish to maintain an ongoing safety net proportional to its loan outstanding $\{L1+L2+L3\}_{i,t}$. This restrictive condition may be relaxed by assuming $\{L1+L2\}_{i,t}$, making interbank credit free from reserve corporate targeting or regulatory requirement.\\
\\
The target reserve parameter $\gamma^{TR}_{i,t}$ captures the bank treasury management policy which aims to maintain an ongoing safety net between cash inflows and holdings, and cash outflows and obligations. This target may include the minimum reserve ratio required by banking law and regulation ($\gamma^{RR}$), the latter being a threshold that is universal across banks and may evolve over time and circumstances.\\
\\
Various definitions of reserve base $\sum{A*_{i,t}}$ involve featuring definitions of the ultimate means of payment and settlement. When A1 only is retained, a narrow monetary system is defined that restricts money functions to currency. When A1 and A3 are jointly retained, a broader monetary system enables banks to settle interbank payments through mutual credit admittances over time and circumstances. When A2 is introduced, a mechanism of loan securitisation is in place, enabling banks to increase their liquidity (defined as held reserve base) through refinancing of their asset holdings. In the rest of this article, we maintain a broad definition of reserve base including A1 and A3 (broad monetary system).\\
\\
The above equation compares with the textbook representation for fractional reserve lending of loanable funds as follows:\\
\\
\underline{Lending Behavior 2:} Fractional reserve under loanable funds constraint\\
\\
Based on current values of the banks' reserve base ($\sum{A*_{i,t}}$) and total deposit $\{L1+L2+L3\}_{i,t}$, individual banks calculate their own target (required) reserves $TR_{i,t}$ (i.e.: the target/required reserve to deal with potential deposit withdrawals):
	\begin{equation}
	TR_{i,t}=\gamma^{TR}_{i,t}\times (L1_{i,t}+L2_{i,t}+L3_{i,t})
	\end{equation}
where $\gamma^{TR}_{i,t}$ is the target reserve ratio for bank $i$ at time period $t$.  
\\
\\Each bank wishes to lend the share of total currently held reserve base to customers as new retail loans ($\Delta A2_{i,t+1}$) above the target reserve threshold:
	\begin{equation}
	\Delta A2_{i,t+1} = \sum{A*_{i,t}} - TR_{i,t}
	\end{equation}
So the general formula of target lending is: 
\begin{equation} \label{eq:LendingB1E0}
	Potential \Delta A2_{i,t+1} = Max\Bigg\{ \ 0 \ ,\ \ \sum{A*_{i,t}}-{\gamma^{TR}_{i,t}} \cdot (L1_{i,t}+L2_{i,t}+L3_{i,t}) \ \Bigg\}
	\end{equation}		
In particular, the banking theory as pure intermediation of loanable funds imposes that only A1 is included in the reserve base $\sum{A*_{i,t}}$.

\subsubsection{Bank credit realisation}
Both lending strategies denote the ongoing bank potential offer of loans to customers at time period $h\geq t$. Banks’ potential offer of loans is then satisfied according to the following aggregate absorption function by customers as a whole.\\
\\
Hence, the actual granted loan:
	\begin{equation}\label{eq:LendingActual}
	Actual \Delta A2_{i,t+1} = \widetilde{\Theta}_{i,t} \cdot Potential \Delta A2_{i,t+1}
	\end{equation}
where the proportion of potential lending $\widetilde{\Theta}_{i,t}$ is generated randomly from a triangular distribution (dependent on the lower, peak, upper parameters):
	\begin{equation}
	\widetilde{\Theta}_{i,t}(\theta_L,\theta_P,\theta_U) \sim Triangular(\theta_L,\theta_P,\theta_U)
	\end{equation} 
This statistical distribution implies an average loan absorption of $(\psi_L+\psi_P+\psi_U)/3$ and a modal loan absorption of $\psi_P$ at each period. These parameters capture business cycle conditions in the lending activity over time and circumstances. For sake of simplicity and without generality loss, we treat customers as a whole, disregarding individual customer patterns of bank loan granted, spent and repaid. A further extension may introduce a set of minimal institutions to design and implement real economy uses and fate of bank credit money.\\
\\
Once each bank $i$ gets its actual share of granted loans, it allocates them to its customers j, increasing simultaneously both bank loan asset portfolio $(A2_{j,h})$ and bank customer deposits $(L2_{j,h})$.

\subsection{Model Module 3: Interbank settlement and credit system}
Both payments settlement (module 1) over time periods t, and bank credit creation and destruction (module 2) over time periods h reshape bank outstanding asset holdings and liability obligations throughout time and space. This ongoing dynamic requires the bank treasury department to manage the bank financial process to rebalance cash (and cash equivalents) outflows and inflows, seeking to maintain ongoing bank capacity to settle obligations when they become due in time and amount. As for banks hold reserves to settle payments and meet reserve requirements \citep{fullwiler2008modern}. This ongoing rebalancing activity is performed through interbank borrowing and lending. When a bank lends to another bank, bank balance sheet simultaneously expands with the loan (an asset to the bank) and the deposit (a liability to the bank), both in the name of the other bank, while the bank promises to make the borrowing bank deposit redeemable in cash or cash equivalents (that is, whatever asset $A*$ is included in reserve base definition $\sum{A*}$). 
 
\subsubsection{Interbank credit repayment system}
Before seeking for new interbank lending and borrowing, all the banks repay their outstanding loans L3 that have become due in period t according to the following steps:
\begin{description}
	\item[Step 1:] Each bank records its outstanding interbank lending and borrowing over time, so every loan has one-to-one match between a lender and a borrower at a particular time t;   
	\item[Step 2:] At each time period, some randomly selected loans come to be repaid (selected from all outstanding loans through all historical periods), depending on the interbank repayment likelihood parameter $\omega_{i,t}$; 
	\item[Step 3:] The control parameter $\omega_{i,t}$ for each bank i at time period t manages the likelihood of repayment and may vary over time and circumstances;
	\item[Step 4:] The repayment is settled by the repaying bank through a proportional transfer of its reserve base components $A*$: for the lender, it decreases its interbank lending item $A3$ and increases its reserve base $A*$; for the borrower, it decreases its interbank borrowing item $L3$ and decreases its reserve bases $A*$;
	\item[Step 5:] After interbank loan repayment, each bank updates its record of outstanding interbank lending and borrowing. 
\end{description}
This simple repayment mechanism is sufficient to denote interbank lending dynamics. A further extension of the model may transform this mechanism in a more sophisticated treasury management function by each bank, and more sophisticated institutional arrangements to frame this management over time and circumstances.\\
\\
The model implements interbank repayment as follows.\\
\\
A 3-dimension matrix $M_{B\times B\times T}$ records all outstanding interbank loans (created through customers' bank credit wire transfer and credit coordination in interbank networking). A remaining loan is recorded as $m_{u\rightarrow v,t}>0, \forall m\in M_{B\times B\times T}$, otherwise if $m_{u\rightarrow v,t}=0$ loan is absent). At each time, another random 3-dimension matrix $\widetilde{D}_{B\times B\times T}$ is generated where each element $\tilde{d}_{u,v,t}$ is drawn from a uniform distribution $\sim U(0,1)$.\\
	\\
	The repayment triggering threshold in the repayment decision matrix $\widetilde{D}_{B\times B\times T}$ is fixed as follows,
	\begin{equation}
	\widetilde{d}_{u,v,t}>\omega_{u,t}, \ \ \ \forall (u,v)\in \{m_{u\rightarrow v,t}>0\}
	\end{equation} 
(where $\omega_{u,t}$ parameter controls for interbank repayment probability, i.e. a higher $\omega_{u,t}$ means that interbank loans are repaid more slowly implying a longer duration on average, and $(u,v)$ identifies a pair of lender-borrower.) \\
\\
When the repayment is triggered, bank $u$ repays interbank loan $m_{u\rightarrow v,t}$ where $\tilde{d}_{u,v,t}>\omega_{u,t}$, therefore $m_{u\rightarrow v,t}$ becomes zero, and the repayment is paid back to the previous lending bank $v$ and adds to its reserve base $\sum{A*_{i,t}}$ in proportion to reserve base components. Borrower reserve accounts $A*$ decrease and may become zero for A1 and negative for other $A*$ because of this repayment. Indeed the repaying bank may seek to recover from this exposure through further inter-bank borrowing.\\
\\
The balance sheet items (A3/L3/A*) are updated as follows:
\begin{itemize}
	\item Once a pair of lender-borrower $(u,v)$ is selected to repay the remaining loan $m_{u\rightarrow v,\tau}$, the lender u's $A3_{u,t}$ and the borrower v's $L3_{v,t}$ at the current period are reduced by each $m_{u\rightarrow v,\tau}, \forall \tau \in \{1...t\} $.
	\item Each pair $(u,v,\tau)$ also identifies the weights of different reserve base component at time $\tau$ as $A*_{i,\tau}/\sum{A*_{i,\tau}}$, so the lender u's each reserve base component $A*_{u,t}$ is increased, while the borrower v's each reserve base component $A*_{v,t}$ is reduced in proportion to reserve components at current or past period levels. 
	\item This process is repeated for all selected remaining loans $m_{u,v,\tau}, \forall u,v,\tau$.
	%\item This process is repeated for all selected remaining loans $m_{u,v,t}, \forall u,v,t$. 
	%The following matrix representation applies:
	%\begin{equation}
	%\begin{array}{l}
  %\Delta A3_{u,t} = -\sum^t_{\tau=1} \sum_{v\in (u,v)} m_{u\rightarrow v,\tau}, \ \ \ \forall (u,v)\in \{\widetilde{d}_{u,v,\tau}>\omega\} \\
	%\Delta A*_{u,t} = +\sum^t_{\tau=1} \sum_{v\in (u,v)} \frac{A*_{u,\tau}}{\sum{A*_{u,\tau}}} \cdot m_{u\rightarrow v,\tau}, \ \ \ \forall (u,v)\in \{\widetilde{d}_{u,v,\tau}>\omega\}  
	%\end{array} 
	%\end{equation}
	%\begin{equation}
	%\begin{array}{l}	
	%\Delta L3_{v,t} = -\sum^t_{\tau=1} \sum_{u\in (u,v)} m_{u\rightarrow v,\tau}, \ \ \ \forall (u,v)\in \{\widetilde{d}_{u,v,\tau}>\omega\} \\
	%\Delta A*_{v,t} = -\sum^t_{\tau=1} \sum_{u\in (u,v)} \frac{A*_{v,\tau}}{\sum{A*_{v,\tau}}} \cdot m_{u\rightarrow v,\tau}, \ \ \ \forall (u,v)\in \{\widetilde{d}_{u,v,\tau}>\omega\} 
  %\end{array} 
	%\end{equation}	
\end{itemize}
After interbank repayment settlements, banks with outstanding excess reserve holdings may seek to lend, while banks with outstanding reserve needs do seek to borrow. In this context, a minimal reserve requirement implies a balance threshold higher than zero to distinguish between potential bank lenders and borrowers.
	
\subsubsection{Interbank credit pooling system}
This section introduces the dynamic interactions of interbank credit coordination. Both the completeness and perfection of this pooling system is managed by a credit control parameter $\phi\in [0,1]$. In particular, with $\phi=0$ the system works under complete and perfect pooling, while with $\phi=1$ the system does not experience any interbank credit. A system is complete when all banks can potentially become borrowers and lenders. A system is perfect when all banks do actually become either borrowers or lenders. In section 4, simulation analysis will comparatively assess the system working against different levels of incomplete and imperfect coordination modes, inferring systemic implications of interbank coordination for financial stability and resilience. The inter-bank coordination quality parameter $\phi_{i,t}$ may be settled at different values across banks $i$ and time periods $t$. For sake of simplicity, our simulation analysis denotes it through one universal value for all $i,t$.\\
\\
Interbank credit coordination mechanism is implemented as follows:
\begin{description}
	\item[Calculating reserve situation:] Based on its individual bank target reserve ratio $\gamma^{TR}_{i,t}$ and its own current reserve base level ($\sum{A*_{i,t}}$), each bank identifies its reserve surplus (denoting a potential lender) or shortage (denoting a potential borrower). \\
	\\
	The current reserve ratio of a bank i (all balance sheet items being updated up to this step) is:
	\begin{equation}
	\gamma^{CR}_{i,t}=\frac{\sum{A*_{i,t}}}{L1_{i,t}+L2_{i,t}+L3_{i,t}}
	\end{equation}
The bank's excess reserve (surplus) when $\gamma^{CR}_{i,t}>\gamma^{TR}_{i,t}$ is:
	\begin{equation}
	ER_{i,t}=(\gamma^{CR}_{i,t}-\gamma^{TR}_{i,t})\cdot(L1_{i,t}+L2_{i,t}+L3_{i,t})
	\end{equation}		
The bank's reserve need (shortage) when $\gamma^{CR}_{i,t}<\gamma^{TR}_{i,t}$ is:
	\begin{equation}
	RN_{i,t}=(\gamma^{TR}_{i,t}-\gamma^{CR}_{i,t})\cdot(L1_{i,t}+L2_{i,t}+L3_{i,t})
	\end{equation}	
	\item[Calculating weights of reserve bases:] since the balance sheet comprises different reserve base components $A*_{i,t}\in (A1_{i,t},A2_{i,t},A3_{i,t})$ which may be used for interbank settlement proportionally to their relative weight in $\sum{A*_{i,t}}$, we compute and record the share of each reserve base component in the total reserve for each bank. When issuing or repaying the interbank loan, the reserve base components will be updated according to these weight:  
	\begin{equation}
  Weight_{A*_{i,t}}=\frac{A*_{i,t}}{\sum{A*_{i,t}}}
 	\end{equation}
	\item[Generating a potential lender-borrower matrix:] All banks are now separated into two groups (i.e. lenders and borrowers). The system forms a matrix $M_{B\times B}$ ($B$ is the total number of banks in the system), lending banks being marked as $1$ into the rows, and those borrowing banks being marked as $1$ into the columns. When a lending bank finds a borrowing bank, their cells in the matrix have both column and row as 1. Hence, this matrix defines all combinations of potential lenders and borrowers. When a lending bank $l$ matches a borrowing bank b, then the element $m_{l\rightarrow b}=1, m\in M_{B\times B}$, defining a pair $(l,b)$. A bank with excess reserve is a potential lender for all borrowers (who need reserve), and vice-versa. Therefore, 
	\begin{equation}
	m_{l\rightarrow b}=1, \ \ \forall l\in B\{\gamma^{CR}_{i,t}>\gamma^{TR}_{i,t} \} \cap \forall b\in B\{\gamma^{CR}_{i,t}<\gamma^{TR}_{i,t} \}
	\end{equation}
	\item[Generating an actual lender-borrower matrix:] Another $B\times B$ matrix is used to define the actual pair of lender-borrower. The system creates a random probability selection matrix $\widetilde{M}^{prob}_{B\times B}$, whose element $\tilde{m}^p_{l,b}$ is generated by one of the following approaches: ``endogenous partner search'' or ``exogenous random matching'' \citep{DelliGatti20101627}. \\
	\\
	\textsl{``Endogenous partner search''}: A preferential attachment parameter is generated for each pair $(l,b)$ according to the lending bank gearing ratio and the borrowing bank relative exposure to interbank credit as follows:
	\begin{equation}
  r_{l\rightarrow b,t}=\alpha (L_{l,t})^{-\alpha}+\alpha (l_{b,t})^{\alpha}, \ \ \ \ \mbox{with } \alpha>0
	\end{equation}
where $L_{l,t}=max\{0; L4_{l,t}/(L1_{l,t}+L2_{l,t}+L3_{l,t}+L5_{l,t})\}$ (i.e. equity ratio over total liability of bank $l$) and $l_{b,t}=L3_{b,t}/(L1_{b,t}+L2_{b,t}+L3_{b,t}+L5_{b,t})$ (i.e. interbank borrowing ratio over total liability of bank $b$) at time $t$. Then, we can follow the partner selection criterion by \citet{DelliGatti20101627} as follows:
\begin{equation}
\tilde{m}^p_{l,b} = \lambda e^{-\lambda \cdot r_{l\rightarrow b,t}}, \ \ \ \forall (l,b)\in \{m_{l\rightarrow b}=1\} \ \ \ \ \mbox{with } \lambda>0
\end{equation}
Comparing with \citet{DelliGatti20101627}, our model has a few differences: (1) our model extends the loan repayment duration to multiple periods (via a stochastic loan repayment ratio $\omega_{i,t}$); (2) our model extends to multi-lenders with multi-borrowers where each lender has a limited funding; (3) they analyze the impact of lending (revenue: interest, or loss: bad debt) to bank profit and loss statement, while we focus on the impact of loan transactions on bank balance sheet dynamics. \\
\\
	\textit{``Exogenous random matching''}: This approach draws the actual matching matrix elements from uniform distribution as follows:
	\begin{equation}
  \tilde{m}^p_{l,b} \sim U(0,1)
	\end{equation}	
	The rest of the procedure is identical for both approaches. Once we have all probabilities for all potential pairs of lender-borrower $(l,b)$, we select those pairs whose probabilities are larger than a threshold to form actual subsets of interbank networks:
	\begin{equation}
  \tilde{m}^p_{l,b} > \phi_t, \ \ \ \ \mbox{with } \phi_t \in [0,1]
	\end{equation}	
	We control this threshold parameter in order to manage the degree of imperfect pooling (i.e.: if $\phi_t=0$, then the network realises perfect pooling). This implementation helps to select multi-lenders with limited funding, contrary to \citet{DelliGatti20101627} which introduce only one lender with unlimited funding available.
	\item[Allocating reserve within subsets of interbank network:] The actual selection matrix may contain either borrowers (lenders) who couldn't find lenders (borrowers), or borrowers (lenders) who have one or more lenders (borrowers). Each borrower decides to borrow in proportion to its reserve shortage from the group of its actual lenders based on those lenders' total excess reserve. Borrowers request the needed amount of reserve from their actual lenders. Each lender forms a subgroup between itself and its actual borrowers. This means that borrowers and lenders form subsets of interbank credit networks where banks coordinate their reserve need and excess for the period $t$ being. Since interbank loans may last for some periods, this makes interbank credit networks evolving over time and circumstances captured by the parameter space. 
	\item[Updating bank balance sheet items:] If a pair of lender-borrower $(l,b)$ performs interbank credit for an amount $\Delta A3_{l\rightarrow b,t}=\Delta L3_{l\rightarrow b,t}$, then the lender's $A3_{l,t}$ and borrower's $L3_{b,t}$ are increased by this amount. This operation is repeated for each coupled transaction. Borrowers and lenders constitute interbank pooling networks. 
	%(We may remove these lines if we are not sure that we maintain this mechanism in the following versions. If you prefer keeping them, please make sure that they respect the logic we agreed together (full repayment of the loan, the same amount being transferred in proportion))Therefore, the lenders' defined reserve base components $A*_{l,t}$ are reduced in proportion to the weights $\frac{A*_{l,t}}{\sum{A*_{l,t}}}$ and the borrowers' are increased in proportion to the weights $\frac{A*_{b,t}}{\sum{A*_{b,t}}}$. For all pairs within a subset, the ones total amounts are:
	%\begin{equation}
	%\begin{array}{ll}
  %\Delta A3_{l,t} = \sum_{b\in (l,b)} \Delta L3_{l,b,t}, & \forall (l,b)\in \{\tilde{m}^p_{l,b} > \phi\} \\
	%\Delta A*_{l,t} = -\sum_{b\in (l,b)} \frac{A*_{l,t}}{\sum{A*_{l,t}}} \cdot \Delta L3_{l,b,t}, & \forall (l,b)\in \{\tilde{m}^p_{l,b} > \phi\} 
	%\end{array} 
	%\end{equation}
	%\begin{equation}
	%\begin{array}{ll}	
  %\Delta L3_{b,t} = \sum_{l\in (l,b)} \Delta A3_{l,b,t}, & \forall (l,b)\in \{\tilde{m}^p_{l,b} > \phi\}	\\
	%\Delta A*_{b,t} = \sum_{l\in (l,b)} \frac{A*_{b,t}}{\sum{A*_{b,t}}} \cdot \Delta A3_{l,b,t}, & \forall (l,b)\in \{\tilde{m}^p_{l,b} > \phi\}  
  %\end{array} 
	%\end{equation}
	\item[Recording interbank loans in the repayment matrix:] Actual new interbank loan transactions at this period $t$ between lenders and borrowers are recorded in the 3-Dimension matrix $M_{B\times B\times T}$, together with all the other outstanding interbank credits and debts. If the interbank loan need is not fully satisfied within a subgroup, the borrowers ask the central bank for overdraft assistance.
\end{description}

\subsection{Central bank as guarantor of last resort}
In some circumstances, the model dynamic makes possible that, after interbank credit coordination, one bank remains exposed, that is, it is still unable to match its reserve base targets and obligations in that time period $t$.\\
\\
In this situation, the model introduces a non-cash central bank facility that acts as a residual guarantee, enabling the exposed bank to roll over its overdraft obligations until the next period $t+1$. For sake of simplicity, this facility is non-cash based, that is, it does not enter the reserve base dotation of the exposed bank, which must pay a punitive guarantee fee on this contingent liability $A5_{i,t}$ outstanding. This modeling strategy is consistent with overdraft facilities provided by major central banks \citep{fullwiler2008modern}. This simplification helps focusing the model dynamic on the impact of interbank credit that is under examination in this article. Central bank exposure may be considered as a signal for financial instability and fragility generated by both bank excess behaviours and interbank lacks of coordination. A further extension of the model may include a set of minimal institutions to design and implement central bank monetary interventions and monetary financial institution defaults.\\
\\
If a bank remains in reserve shortage after interbank credit coordination, that is, $\sum{A*}$ is less than target reserve $TR_{i,t}$, the central bank provides a guarantee $L5_{i,t}$ to complete the level of reserves with:
\[A5_{i,t}=L5_{i,t}\]
Therefore:
	\begin{equation}
  \sum{A*_{i,t}}+A5_{i,t} = TR_{i,t}
	\end{equation}
The assisted bank pays an overdraft guarantee fee on the central bank assistance, as follows:
	\begin{equation}
  \pi_{i,t}=\pi_{i,t-1}-r^{L5}_{i,t}\cdot L5_{i,t}
	\end{equation}
where $r^{L5}_{i,t}$ is equal to the interbank rate $r^{L3}_{i,t}$ plus punitive spread. \\
\\
This assistance from the central bank is fully removed at the beginning of the next period $t+1$ as follows:
	\begin{equation}
	\Delta A5_{i,t+1} = \Delta L5_{i,t+1} = -L5_{i,t}
	\end{equation}

\subsection{Bank Equity dynamics}
Bank equity was initialized by non-cash provision equally distributed across banks. We further model equity dynamics period through period. Following \citet{Xu2016131}, fee and interest rates of reference fluctuate according to exogenous stochastic distributions. A further extension of the model may include a set of minimal institutions to design and implement interdependency between these rates and underlying banking and financial conditions and circumstances.\\
\\
Once bank balance sheets have been updated by all the four steps (payments settlement; bank credit creation and destruction; interbank settlement and credit; central bank guarantee), each bank accrues non-cash movements to its income statement as follows:
\begin{description}
	\item[$\diamond$] Concerning bank lending department with customers:
	\begin{itemize}
		\item Borrowing interest rates pay to customers over outstanding currency deposit $L1$
		\item Management fees as expense to customers over held currency deposits $A1$
		\item Borrowing interest rates pay to customers over outstanding currency deposit $L2$
		\item Lending interest charges as revenues from customers over outstanding loans $A2$
	\end{itemize}
	\item[$\diamond$] Concerning interbank lending and borrowing:
	\begin{itemize}
		\item Interbank lending interest charges as revenues from borrowing banks over outstanding interbank credits $A3$. Conversely, the same amounts become expenses for borrowing banks. 
		\item Interbank borrowing interest charges as expenses to lending banks over outstanding loans $L3$. Conversely, the same amounts become revenues for lending banks.
	\end{itemize}
	\item[$\diamond$] Guarantee fees as expenses incurred by the guaranteed bank over temporary central bank assistance $L5$
\end{description}
Bank equity is updated as follows:
\begin{equation} \label{eq:Figure8Equ1}
L4_{i,t} = A4_{i,t} = L4_{i,t-1} + \pi_{i,t}
\end{equation} 
where profit $\pi_{i,t}$ is the sum of all these movements:
\begin{equation} \label{eq:Figure8Equ2}
\pi_{i,t}=(r^{A1}_{i,t}A1_{i,t}+r^{A2}_{i,t}A2_{i,t}+r^{A3}_{i,t}A3_{i,t})-(r^{L1}_{i,t}L1_{i,t}+r^{L2}_{i,t}L2_{i,t}+r^{L3}_{i,t}L3_{i,t}+r^{L5}_{i,t}L5_{i,t})
\end{equation}
where $r^{A3}_{i,t}$ and $r^{L3}_{i,t}$ are equal for each couple between lender and borrower. \\
\\
As bank balance sheet amounts evolve according to their own movements over time and circumstances, this equity dynamic enables the model to capture the ongoing impact of various bank activities on bank profits and losses. For sake of simplicity, this dynamics is non-cash based, that is, it does not enter the reserve base dotation of the bank period through period and it is not distributed to management and shareholders. This simplification helps focusing the model dynamic on the impact of interbank credit that is under examination in this article. Bank Equity dynamic may be considered as a signal for financial resilience, showing the ongoing capacity of each bank and the banking system as a whole to face possible credit losses. A further extension of the model may include a set of minimal institutions to design and implement cash equity transactions, including transactions with shareholders and bank default resolutions. \\
\\
The Table~\ref{table:BaseCaseParameters} summarizes key parameters, and their benchmark values and features.

\begin{table}[htbp]
%\footnotesize
\scriptsize
\caption{Summary of parameters and values in the model} % title of Table
\centering % used for centering table
\begin{tabular}{|l l l l l|} % centered columns (3 columns)
\hline\hline %inserts double horizontal lines
Parameter & Description & Benchmark & Variation & Baseline Calibration (Section 4)\\ [0.5ex] % inserts table
%heading
\hline % inserts single horizontal line
T & Total time periods &  & Fixed  &  50  \\
B & Total number of banks &  & Fixed &  10 \\
C & Total number of customers &   & Fixed &  1000 \\
         &       &     &  & \\
$A1_0$ & Total base money & $1\times 10^9$ & Fixed &  $1\times 10^9$ \\
$A4_0$ & Total capital of all banks & $1\times 10^8$ & Fixed & $1\times 10^8$\\
         &       &     &  & \\								
$r^{A1}_{i,t}$ & Interest rate of asset A1 & 0.01 & [0.005, 0.015] & $\sim Triangular(0.005,0.01,0.015)$ \\
$r^{A2}_{i,t}$ & Interest rate of asset A2 & 0.03 & [0.02, 0.04] & $\sim Triangular(0.02,0.03,0.04)$\\
$r^{A3}_{i,t}$ & Interest rate of asset A3 & 0.015 & [0.005, 0.025] & $\sim Triangular(0.005,0.015,0.025)$ \\
$r^{L1}_{i,t}$ & Interest rate of liability L1 & 0.01 & [0.005, 0.015] & $\sim Triangular(0.005,0.01,0.015)$ \\
$r^{L2}_{i,t}$ & Interest rate of liability L2 & 0.01 & [0.005, 0.015] & $\sim Triangular(0.005,0.01,0.015)$ \\
$r^{L3}_{i,t}$ & Interest rate of liability L3 & 0.015 & [0.005, 0.025] & $\sim Triangular(0.005,0.015,0.025)$\\
$r^{L5}_{i,t}$ & Interest rate of liability L5 & $r^{L3}_{i,t}+0.03$ & [0.035, 0.055] & $r^{L3}_{i,t}+0.03$ \\
         &       &     &  & \\	
$\gamma^{RR}$ & Bank required reserve ratio & 0.1 & Fixed &  0.1 \\			
$\gamma^{TR}_{i,t}$ & Bank target reserve ratio & $\gamma^{RR}+\tilde{z}$ & Stochastic & $\gamma^{RR}$\\
         &       &     &   &\\	
$\alpha$ & Partner selection & not used & Fixed &  n/a in ``Exogenous matching''\\
$\lambda$ & Partner selection & not used & Fixed &  n/a in ``Exogenous matching'' \\
$\phi_{i,t}$ & Interbank credit pooling quality & 0 & [0,1] & 0/0.4/0.8\\
$\theta_L,\theta_P,\theta_U$ & Actual customer loan creation & (0,0.5,1) & [0,1] & $\sim Triangular(0,0.8,1)$ \\
$\psi_L,\psi_P,\psi_U$ & Customer loan repayment & (0,0.5,1) & [0,1] & $\sim Triangular(0,0.3,1)$ \\
$\omega_{i,t}$ & Interbank loan repayment   & (0,0.5,1) & [0,1] & 0.5 \\
$\xi_1$  & Scale of customers' cash payment & 0.1 & [0,1] & 0.1 \\
$\xi_2$ & Scale of bank credit wire transfer & 0.1 & [0,1] & 0.1 \\
         &       &     &  & \\				
\hline %inserts single line
\end{tabular}
\label{table:BaseCaseParameters} % is used to refer this table in the text
\end{table}

\newpage
\section{Interbank credit and the money generation process}
\begin{figure}[htbp]
	\centering	\includegraphics[width=1\textwidth]{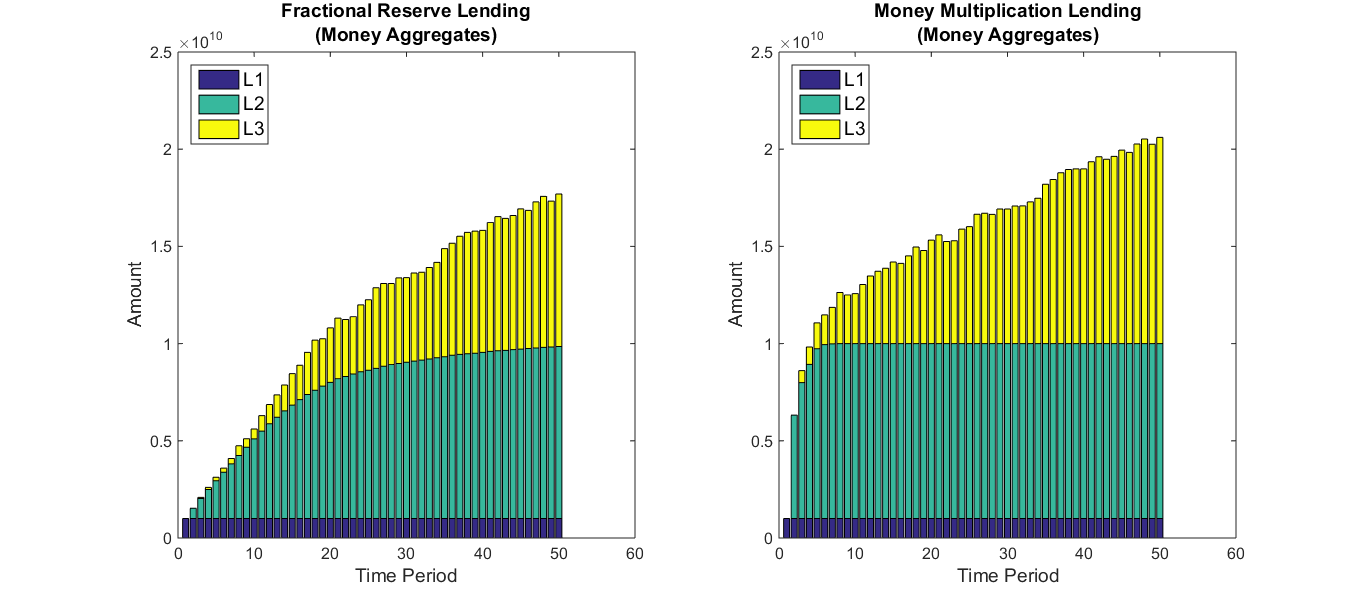}
	\caption{\footnotesize{\textit{Aggregate analysis of narrow banking: both visualizations are generated based on the benchmark values in Table~\ref{table:BaseCaseParameters}, when the interbank coordination is in perfect pooling, all banks use $A1$ as the reserve base (i.e. narrow banking), no repayment of customer loans, the interbank loans are repaid randomly and on average repaid fully in two time periods. The only difference is: the left diagram shows the simulation result when all banks' lending strategy is based on the conventional rule of fractional reserve, while the right diagram illustrates the money multiplication lending strategy.}}}
	\label{fig:Figure1Section3}
\end{figure}

\begin{figure}[htbp]
	\centering	\includegraphics[width=1\textwidth]{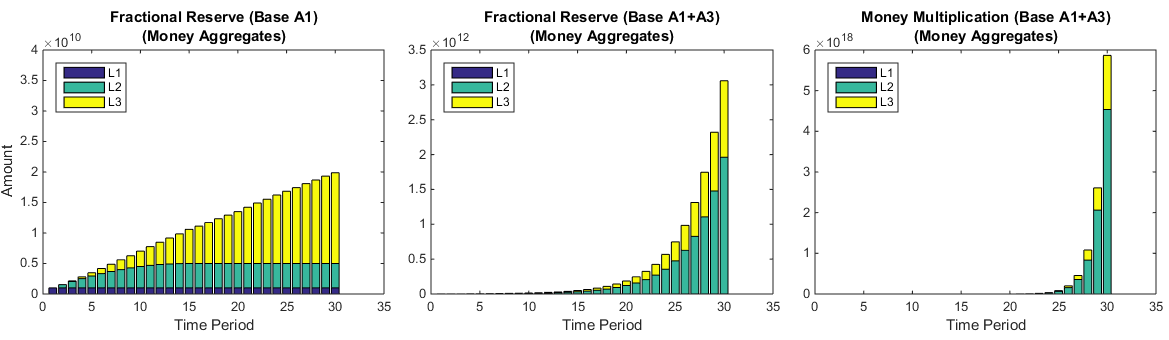}
	\caption{\footnotesize{\textit{Aggregate analysis of broad banking: all three visualizations are generated based on the benchmark values in Table~\ref{table:BaseCaseParameters}, when the interbank coordination is in perfect pooling, no repayment of customer loans, but also no repayment of interbank loans. However, the left diagram represents the simulation result when all banks use $A1$ as the reserve base (i.e. narrow banking), while banks in both the middle and right diagrams use $A1+A3$ as reserve base (i.e. broad banking). Another difference is: banks in both the left and middle diagrams use the lending strategy of fractional reserve, while the right diagram is based on the lending strategy of money multiplication.}}}
	\label{fig:Figure2Section3}
\end{figure}

Our model features the connection between money as a means of payment and bank credit generation. In particular, banks define their lending strategy in relation to their ongoing capacity to face payment settlement obligations over time and circumstances. This implies that banks watch over ongoing inflows and outflows of means of payment (cash and cash equivalents) in order to target their ongoing potential capacity to offer loans period through period. In this context, bank reserve holdings are defined through whatever bank assets $A*$ that may be used to settle payments with other banks. These assets jointly define the reserve base that the bank takes into account to decide its lending capacity through leverage, enabling its contribution to the money multiplication process. \citet[pp. 320]{Schumpeter1954} masterly denotes this process as money manufacturing. In this context, ``banks are no longer said to `lend their deposits' or `other people's money,' but to `create' deposits or bank notes: they appear to manufacture money rather than to increase its velocity or to act, which is a completely unrealistic idea, on behalf of their depositors.''\\
\\
According to textbook story of money creation under fractional reserves and available loanable funds, interbank credit is relegated to the backyard of the banking system. It may be considered as a minor mechanism that enables banks to temporarily postpone payments in order to facilitate settlement and clearing across banks, with no impact on the overall working of the banking system. Our model is capable to reproduce this narrow monetary system as a cornerstone case which occurs when the reserve base is narrowly defined as currency only, that is, $A1$. This case may be reproduced under both lending mechanisms (Equation~\ref{eq:LendingB1E0} and~\ref{eq:LendingB2E0}) which feature fractional reserve and money multiplication processes. Figure~\ref{fig:Figure1Section3} illustrates this result.\\
\\
However, the functional equivalence between currency and deposit introduces an additional source of money creation into the financial system. This source comes from the working banking system which, through interbank credit, makes itself potentially independent from base money issued by central banking. Heuristically, this may happen not only when central banking expands its reserve base definition to admit bank securities and bank credit admittances ($A3_{i,t}$) to its refinancing facilities, but also when banks keep renewing drawing facilities to each other ($L3_{i,t}$), or keep issuing credit admittances ($A3_{i,t}$) that circulate as cash equivalents for settling interbank payments over time. Figure~\ref{fig:Figure2Section3} illustrates this scenario.\\
\\
Under ideal conditions similar to textbook case (Figure~\ref{fig:Figure2Section3} case: left diagram), broad interbank credit mechanism involves two featuring results that contrast with received textbook story. According to the latter, interbank credit does not matter and the aggregate money creation is bound by the outstanding reserve base issued by the central bank (high powered money). In contrast, on the one hand, a banking system embedded into a broad monetary system generates an additional money aggregate made of interbank credit money which complements both bank credit money and currency money aggregates. On the other hand, it makes the potential offer of bank credit potentially unbound (Figure~\ref{fig:Figure2Section3} cases: middle and right diagrams).\\
\\
Together, these featuring results pave the way to a better understanding of systemic implications of interbank credit for financial stability and resilience. As for interbank credit generates additional aggregate money while increasing bank leverage and exposure. It further expands bank credit capacity potentially without limits or, better, within evolving limits endogenously imposed by real economy absorption and institutional arrangements. The next section applies this model to assess the impact on financial stability and resilience of this broad banking system under conditions of incomplete and imperfect interbank credit pooling.

\section{Financial stability and interbank credit coordination modes}
\begin{figure}[htbp]
	\centering	\includegraphics[width=1\textwidth]{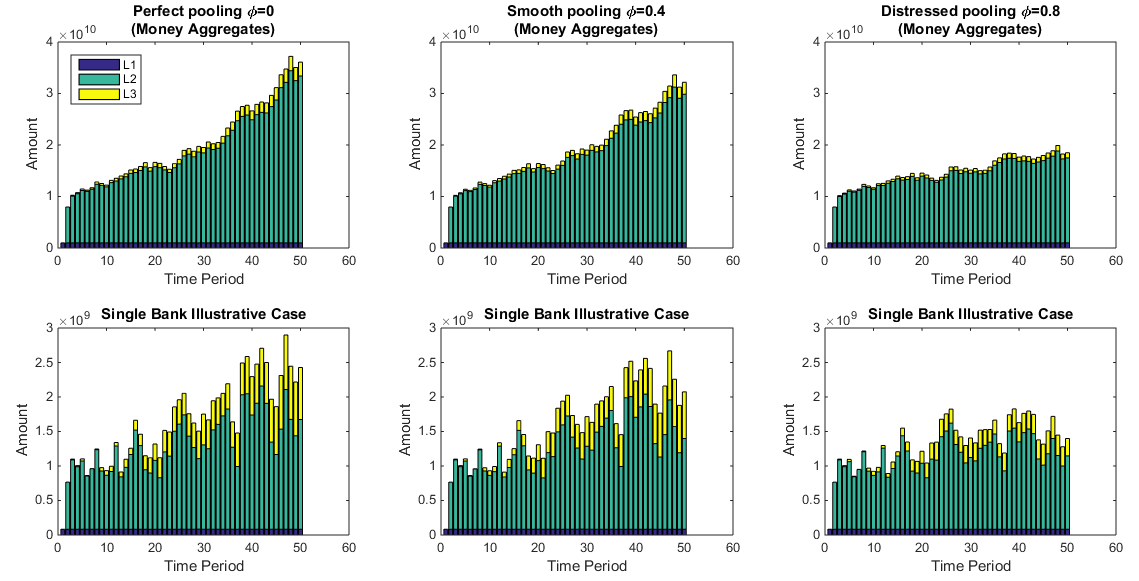}
	\caption{\footnotesize{\textit{Money aggregate dynamics under baseline calibration (parameter values are summarized in the Table~\ref{table:BaseCaseParameters}). Three degrees of inter-bank coordination are analyzed (i.e. perfect pooling $\phi=0$, smooth pooling $\phi=0.4$, and distressed pooling $\phi=0.8$). Panel 1 shows the total amounts of money aggregates in the systems of three analyzed cases, and Panel 2 provides a single bank illustrative case.}}}
	\label{fig:Figure3Section4}
\end{figure}

\begin{figure}[htbp]
	\centering	\includegraphics[width=1\textwidth]{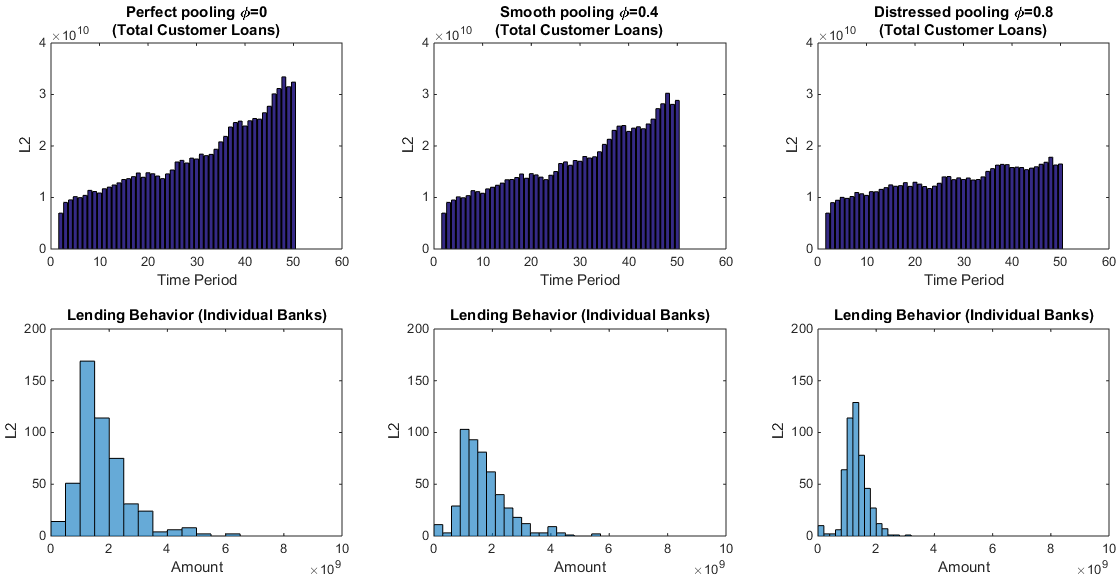}
	\caption{\footnotesize{\textit{Customer lending under baseline calibration (parameter values are summarized in the Table~\ref{table:BaseCaseParameters}). Three degrees of inter-bank coordination are analyzed (i.e. perfect pooling $\phi=0$, smooth pooling $\phi=0.4$, and distressed pooling $\phi=0.8$). Panel 1 shows the total accumulated amounts of customer loans in the systems of three analyzed cases over time, and Panel 2 provides the histograms of all individual banks' customer loans.}}}
	\label{fig:Figure4Section4}
\end{figure}

\begin{figure}[htbp]
	\centering	\includegraphics[width=1\textwidth]{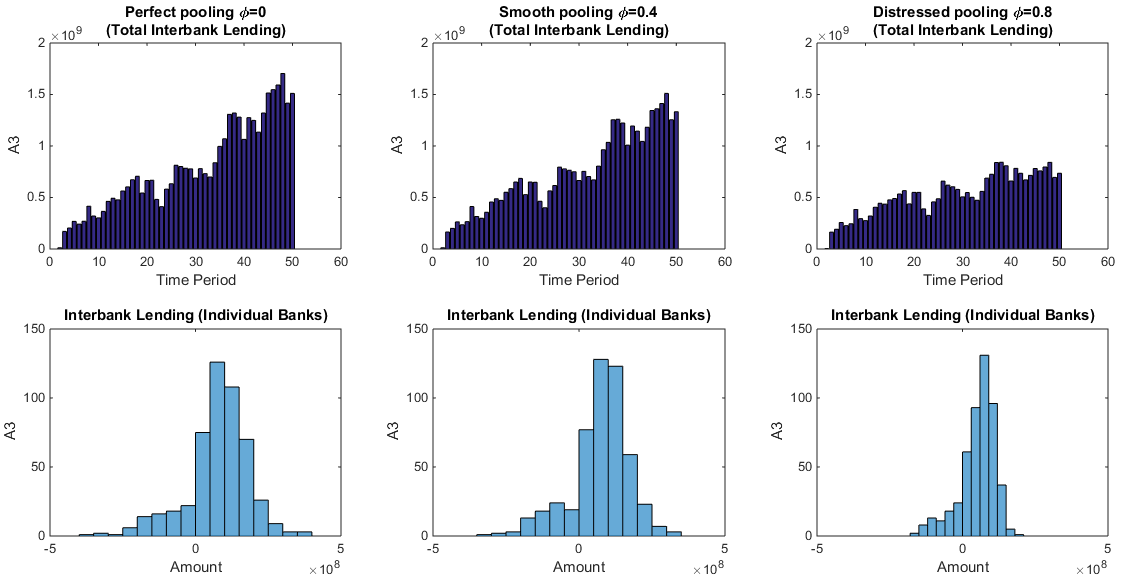}
	\caption{\footnotesize{\textit{Interbank lending under baseline calibration (parameter values are summarized in the Table~\ref{table:BaseCaseParameters}). Three degrees of inter-bank coordination are analyzed (i.e. perfect pooling $\phi=0$, smooth pooling $\phi=0.4$, and distressed pooling $\phi=0.8$). Panel 1 shows the total accumulated amounts of interbank lending in the system of three analyzed cases over time, and Panel 2 provides the histograms of all individual banks' interbank lending.}}}
	\label{fig:Figure5Section4}
\end{figure}

\begin{figure}[htbp]
	\centering	\includegraphics[width=1\textwidth]{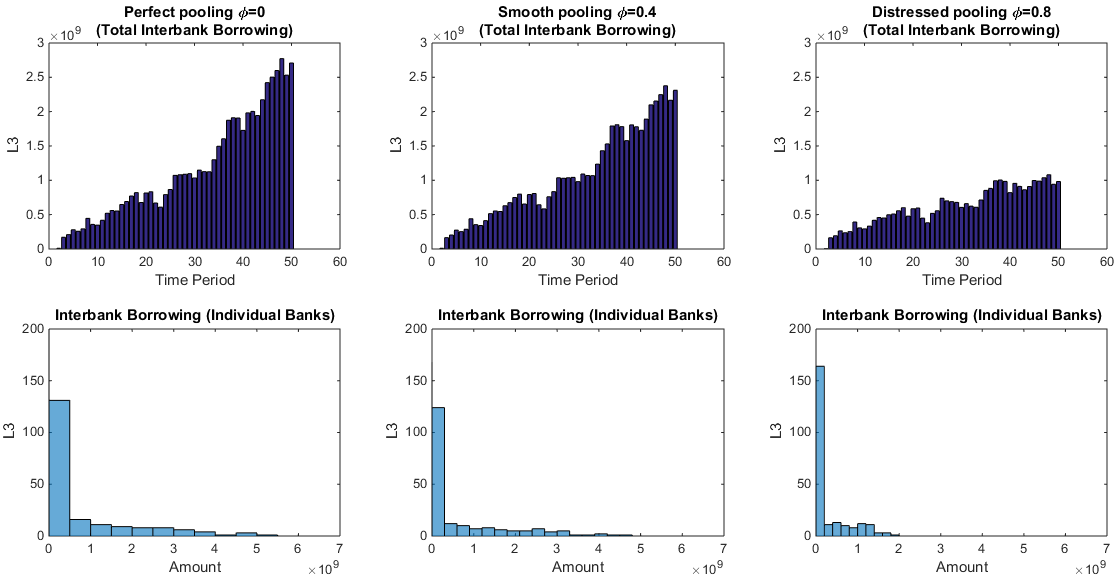}
	\caption{\footnotesize{\textit{Interbank borrowing under baseline calibration (parameter values are summarized in the Table~\ref{table:BaseCaseParameters}). Three degrees of inter-bank coordination are analyzed (i.e. perfect pooling $\phi=0$, smooth pooling $\phi=0.4$, and distressed pooling $\phi=0.8$). Panel 1 shows the total accumulated amounts of interbank borrowing in the system of three analyzed cases over time, and Panel 2 provides the histograms of all individual banks' interbank borrowing.}}}
	\label{fig:Figure6Section4}
\end{figure}

\begin{figure}[htbp]
	\centering	\includegraphics[width=1\textwidth]{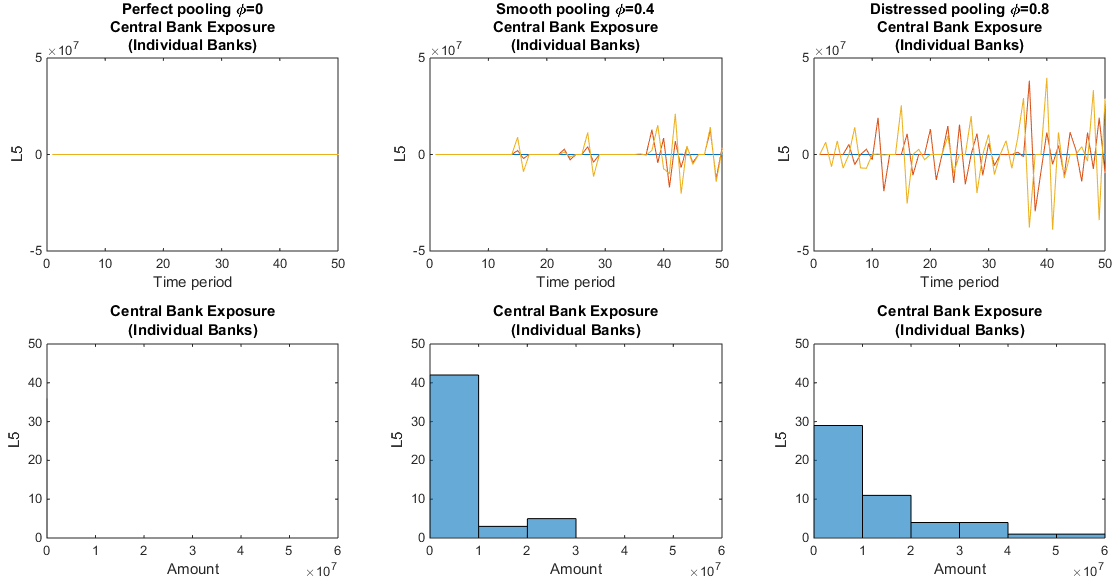}
	\caption{\footnotesize{\textit{Central bank assistance under baseline calibration (parameter values are summarized in the Table~\ref{table:BaseCaseParameters}). Three degrees of inter-bank coordination are analyzed (i.e. perfect pooling $\phi=0$, smooth pooling $\phi=0.4$, and distressed pooling $\phi=0.8$). Panel 1 shows the amounts of central bank assistance in the system of three analyzed cases over time, and Panel 2 provides the histograms of all individual central bank assistance.}}}
	\label{fig:Figure7Section4}
\end{figure}

\begin{figure}[htbp]
	\centering	\includegraphics[width=1\textwidth]{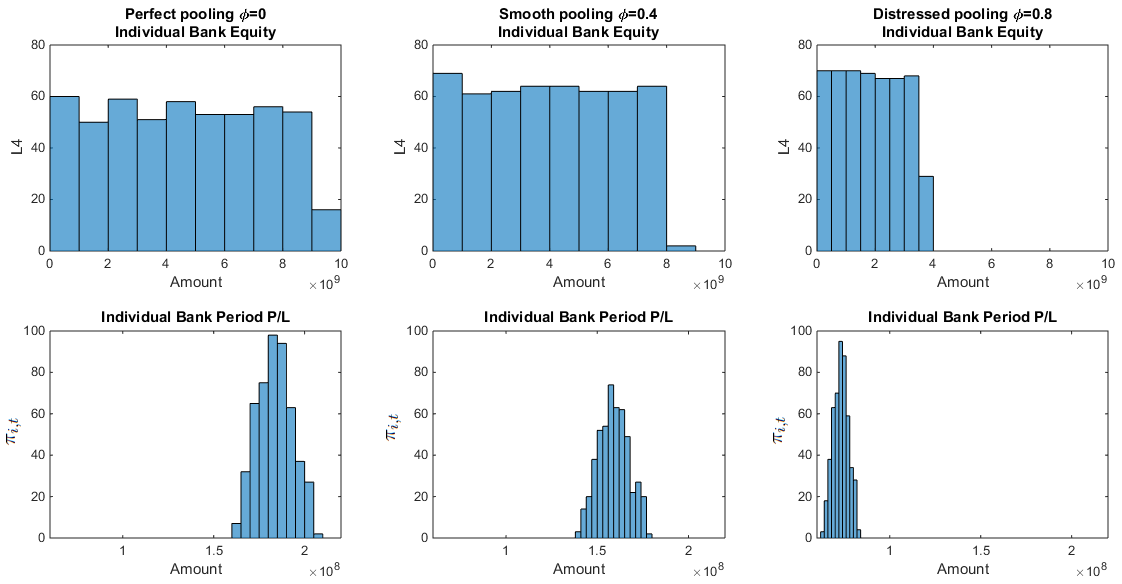}
	\caption{\footnotesize{\textit{Bank equity dynamics over time and circumstances under baseline calibration (parameter values are summarized in the Table~\ref{table:BaseCaseParameters}). Three degrees of inter-bank coordination are analyzed (i.e. perfect pooling $\phi=0$, smooth pooling $\phi=0.4$, and distressed pooling $\phi=0.8$). Panel 1 shows the histograms of bank equity $(L4_{i,t})$ over time in the system of three analyzed cases (Equation~\ref{eq:Figure8Equ1}), and Panel 2 provides the histograms of all individual banks' profit or loss $(\pi_{i,t})$ in each time period (Equation~\ref{eq:Figure8Equ2}).}}}
	\label{fig:Figure8Section4}
\end{figure}

Section 2 introduced our model of bank credit creation featured by interbank credit under the functional equivalence between currency and deposit. Section 3 illustrated the working of the banking system in case of perfect and complete interbank credit pooling across all the monetary financial institutions (banks). \\
\\
Complete and perfect pooling provides a benchmark representation against which we may comparatively assess the impact of interbank credit coordination on financial stability and resilience. This ideal coordination mode enables all the banks to access to all the others at each period (complete pooling network), with all banks being interested in either lending to, or borrowing from each other (perfect pooling network). Therefore, the reserve base that is factually segregated across banks is completely and perfectly pooled among them, in view to assure interbank settlement and credit at each time period $t$.\\
\\
This section 4 develops some visualizations through simulation analysis to compare results under this ideal coordination mode with two alternative modes of incomplete and imperfect pooling. One mode is featured by a smooth degree of inter-bank coordination (with $\phi = 0.4$). Another mode is featured by a distressed degree of inter-bank coordination (with $\phi = 0.8$). When inter-bank credit conditions worsen, it becomes increasingly difficult for potential borrowing banks to find counter-parties willing to lend (captured by decreasing levels of inter-bank coordination quality parameter $\phi$). This comparative analysis may help inferring implications of banking system dynamic for financial stability and resilience.\\
\\
The following simulation analysis applies a baseline calibration that is common to each scenario (Table~\ref{table:BaseCaseParameters}). This calibration sets the same economic and financial conditions through all time periods and all visualizations. Customer and interbank credit repayments are now activated which dynamically offset the ongoing credit creation toward customers and other banks. By assumption, interbank credit is repaid quicker than customer credit. For sake of simplicity, no default occurs on either credit facility. Our simulation analysis focuses on systemic outcomes under featured conditions of interbank coordination. In particular, we visualize results for four systemic outcomes under three levels of interbank credit coordination quality (labeled respectively perfect, smooth, and distressed pooling). The four systemic outcomes point to: (i) customer lending; (ii) interbank credit (lending and borrowing); (iii) central bank recourse; and (iv) bank equity and profit and loss at each period. The following preliminary results may be further developed and corroborated $-$ including through replication, robustness and sensitivity $-$ to study interbank credit impact on financial stability and resilience for each bank and the banking system as a whole.\\
\\
For sake of comparison with visualizations in section 3 (Figures~\ref{fig:Figure1Section3} and~\ref{fig:Figure2Section3}), the money aggregate dynamics under baseline calibration is reproduced in Figure~\ref{fig:Figure3Section4}. A single bank illustrative case (Figure~\ref{fig:Figure3Section4} panel 2) shows that banks follow their own heterogeneous time patterns for each money aggregate $\{L1_{i,t} ; L2_{i,t} ; L3_{i,t} \}$, which sums up to generate a banking system with moderate customer loan growth over time (Figure~\ref{fig:Figure3Section4} panel 1).\\
\\  
Concerning customer lending (Figure~\ref{fig:Figure4Section4}), better interbank credit coordination (perfect and smooth pooling scenarios) involves higher aggregate lending to customers. Moreover, some individual banks can also maintain larger customer loan portfolios over time in this circumstance. This result is qualified by the quite heroic assumption that real economy customers do keep absorbing bank loan offers over time without limit or default. It shall be further tested by replication and sensitivity analysis over the parameter space.\\
\\
Increased customer lending does further involve an increased interdependency on the flow of payments that are performed through bank wire transfer movements across customer bank deposits $L2_{i,t}$. Concerning interbank lending (Figure~\ref{fig:Figure5Section4}) and borrowing (Figure~\ref{fig:Figure6Section4}), individual interbank credit behaves as expected to worsening interbank coordination quality (decreasing level of $\phi$). Worsened conditions reduce interbank lending both in aggregate (Figure~\ref{fig:Figure5Section4} panel 1) and at the level of individual banks (Figure~\ref{fig:Figure5Section4} panel 2). Therefore, borrowing banks are restrained in their access to interbank credit (Figure~\ref{fig:Figure6Section4}), both in aggregate (Figure~\ref{fig:Figure6Section4} panel 1) and for individual amounts (Figure~\ref{fig:Figure6Section4} panel 2).\\
\\ 
Although better interbank credit conditions involve larger and more intense exposure to interbank borrowing (Figure~\ref{fig:Figure6Section4}), this does not generate increased recourse to central bank assistance (Figure~\ref{fig:Figure7Section4}). Deterioration in interbank coordination (smooth and distressed pooling scenarios) generates fragilities in individual and collective bank exposures. As long as interbank coordination works perfectly (perfect pooling scenario), aggregate and individual recourses to central bank assistance are immaterial and then virtually non-existent (Figure~\ref{fig:Figure7Section4}, perfect pooling scenario). When interbank conditions worsen, banks are increasingly forced to have recourse to central bank overdraft facility to keep their affairs ongoing. Their recourse becomes more intense in amount and more frequent in time, involving longer and more acute state of distress for individual banks and then the banking system as a whole. As expected, distressed pooling scenario makes the central bank recourse $L5_{i,t}$ the most frequent in time and the most intense in both aggregate and individual amounts. Moreover, the banking system needs assistance almost permanently under distressed pooling, while assistance requests occur later and sporadically under smooth pooling.\\
\\  
Previous analysis shows that interbank credit coordination has impact on customer loan, interbank borrowing and recourse to central bank assistance. Consequently, it shapes bank equity dynamics over time and circumstances (Figure~\ref{fig:Figure8Section4}). Under lower interbank coordination quality (smooth and distressed pooling scenarios), bank equity ($L4_{i,t}$) remains thinner on average and more concentrated on lower amounts (Figure~\ref{fig:Figure8Section4} panel 1). Bank profits and losses $(P/L)$ of the period $\pi_{i,t}$ (i.e. $\Delta L4_{i,t}$) are also materially lower, due to less revenue from customer loans and increased expense for central bank assistance. Therefore, under lower interbank coordination quality (smooth and distressed pooling scenarios), banks are increasingly unable to accumulate equity reserves through retained earnings, in view to protect themselves against possible losses incurred in ongoing customer and interbank lending. In fact, this result is qualified by the quite heroic assumption that banks do not distribute dividends to management and shareholders. It may yet hold when this distribution remains in line with realized earnings over time and circumstances.

\section{Conclusive remarks}
Troubles in interbank credit coordination may trigger systemic crises. This happened when, since summer 2007, interbank credit did not longer work smoothly across financial institutions, leading to central bank interventions and government bailouts to protect financial stability and assure financial resilience. While this crisis was accidentally triggered by rising actual and expected defaults in some bank asset categories, the origins of the global financial meltdown were found in systemic fragilities related to bank excess behaviours and the overarching dynamics of interbank credit. \\
\\
Our article develops an interacting heterogeneous agents-based model of interbank credit coordination under minimal institutions, taking a systemic perspective on the relationship between bank credit, money creation and interbank credit. Specific attention is paid to the way banks perform credit creation along with interbank credit, clearing, and payment settlements. \\
\\
First, our analysis explores the relationship between interbank credit coordination and aggregate money generation process. Contrary to received understanding, interbank credit has the capacity to make the monetary system unbound in some configurations. Second, a simulation analysis studies modes of interbank credit coordination, addressing interbank dynamics impact on financial stability and resilience at individual and aggregate levels. Systemically destabilizing forces prove to be related to the working of the banking system over time, especially interbank coordination conditions and circumstances.

\newpage 
\begin{appendices}
\chapter{}
%\newpage
\singlespacing
\section{Summary of other notations } 
\begin{itemize}
\singlespacing
	\item $b_{i}$: a bank $i=1...B$, where B denotes the set of all banks in the system
	\item $c_{j}$: a customer $j=1...C$, where C denotes the set of all customers
	\item $M_{c_j\rightarrow b_i}$: a fixed matrix that identifies a customer j's bank i 
	\item $m_{j\rightarrow i}$: each element in $M_{c_j\rightarrow b_i}$, if j's bank is i, then $m_{j,i}=1$, otherwise 0
	\item $\widetilde{M}_{c_x\rightarrow c_y}$: a stochastic matrix that models a customer $x$ pay customer $y$ 
	\item $\tilde{m}_{x\rightarrow y}$: each element in $\widetilde{M}_{c_x\rightarrow c_y}$ that states a percentage of customer $x$'s deposit pay to $y$
	\item $\widetilde{M}_{b_u\rightarrow b_v}$: a stochastic matrix that models a bank $u$ pay $v$ 
	\item $\tilde{m}_{u\rightarrow v}$: each element in $\widetilde{M}_{b_u\rightarrow b_v}$ that states an amount bank $u$ pay bank $v$
	\item $l_{i,j,t}$: a customer j's deposit in its bank i at time t (e.g. $l1_{i,j,t}$ is cash deposit)
	\item $\pi_{i,t}$: a bank i's profit and loss at time t 
	\item $\Delta$: an incremental change
	\item $\widetilde{\Psi}_{i,t}$: a stochastic customer loan repayment ratio 
	\item $\widetilde{\Theta}_{i,t}$: a stochastic proportion of potential target loan that is actually granted
	\item $Triangular(Lower,Peak,Upper)$: a triangular distribution with parameters
	\item $M_{B\times B\times T}$: A 3-dimension matrix records all remaining interbank loans
	\item $m_{u\rightarrow v,t}$: each element in $M_{B\times B\times T}$ that records a remaining interbank loan from bank u to bank v at time t
	\item $\gamma^{CR}_{i,t}$: a bank's current reserve ratio
	\item $ER_{i,t}$: a bank's excess reserve
	\item $RN_{i,t}$: a bank's reserve needs
	\item $TR_{i,t}$: a bank's target reserve
\end{itemize}

\end{appendices}

%%%%%%%%%%%%%%%%%%%%%%%%%%%%%%%%%%%%%%%%%%%%%%%%%%
%%\backmatter
\newpage
%%\singlespacing
%\footnotesize 
\bibliographystyle{apalike}	
\bibliography{ReferenceLabEXReFi2017}	

\begin{thebibliography}{}

\bibitem[Anand et~al., 2013]{Anand2013219}
Anand, K., Gai, P., Kapadia, S., Brennan, S., and Willison, M. (2013).
\newblock A network model of financial system resilience.
\newblock {\em Journal of Economic Behavior \& Organization}, 85:219 -- 235.
\newblock Financial Sector Performance and Risk.

\bibitem[Anand et~al., 2012]{Anand20121088}
Anand, K., Gai, P., and Marsili, M. (2012).
\newblock Rollover risk, network structure and systemic financial crises.
\newblock {\em Journal of Economic Dynamics and Control}, 36(8):1088 -- 1100.
\newblock Quantifying and Understanding Dysfunctions in Financial Markets.

\bibitem[Aymanns and Farmer, 2015]{Aymanns2015155}
Aymanns, C. and Farmer, J.~D. (2015).
\newblock The dynamics of the leverage cycle.
\newblock {\em Journal of Economic Dynamics and Control}, 50:155 -- 179.
\newblock Crises and ComplexityComplexity Research Initiative for Systemic
  InstabilitieS (CRISIS) Workshop 2013.

\bibitem[Bank~of England, 2008]{Bank2008}
Bank~of England, U. (2008).
\newblock Financial stability report.
\newblock {\em Issue No. 24 October}.

\bibitem[Bargigli et~al., 2014]{Bargigli2014109}
Bargigli, L., Gallegati, M., Riccetti, L., and Russo, A. (2014).
\newblock Network analysis and calibration of the â€œleveraged
  network-based financial acceleratorâ€�.
\newblock {\em Journal of Economic Behavior \& Organization}, 99:109 -- 125.

\bibitem[Bargigli and Tedeschi, 2014]{Bargigli20141}
Bargigli, L. and Tedeschi, G. (2014).
\newblock Interaction in agent-based economics: A survey on the network
  approach.
\newblock {\em Physica A: Statistical Mechanics and its Applications}, 399:1 --
  15.

\bibitem[Battiston et~al., 2012]{Battiston20121121}
Battiston, S., Gatti, D.~D., Gallegati, M., Greenwald, B., and Stiglitz, J.~E.
  (2012).
\newblock Liaisons dangereuses: Increasing connectivity, risk sharing, and
  systemic risk.
\newblock {\em Journal of Economic Dynamics and Control}, 36(8):1121 -- 1141.
\newblock Quantifying and Understanding Dysfunctions in Financial Markets.

\bibitem[Bech et~al., 2015]{Bech201544}
Bech, M.~L., Bergstrom, C.~T., Rosvall, M., and Garratt, R.~J. (2015).
\newblock Mapping change in the overnight money market.
\newblock {\em Physica A: Statistical Mechanics and its Applications}, 424:44
  -- 51.

\bibitem[Beck et~al., 2014]{Beck20141}
Beck, T., Colciago, A., and Pfajfar, D. (2014).
\newblock The role of financial intermediaries in monetary policy transmission.
\newblock {\em Journal of Economic Dynamics and Control}, 43:1 -- 11.
\newblock The Role of Financial Intermediaries in Monetary Policy Transmission.

\bibitem[Biondi, 2017]{Yuri2017Banking}
Biondi, Y. (2017).
\newblock Banking, money and credit: A systemic perspective.
\newblock {\em Financial Regulation Research Lab (Labex ReFi) Working Paper
  Paris}.

\bibitem[Black, 1970]{Black1970}
Black, F. (1970).
\newblock Banking and interest rates in a world without money: The effects of
  uncontrolled banking.
\newblock {\em Journal of Bank Research}, 1:8--20.

\bibitem[Blair, 2013]{Blair2013}
Blair, M.~M. (2013).
\newblock Making money: Leverage and private sector money creation.
\newblock {\em Seattle University Law Review}, 36:417--454.

\bibitem[Blanchard, 2009]{Blanchard2009}
Blanchard, O.~J. (2009).
\newblock The state of macro.
\newblock {\em Annual Review of Economics}, 1:209--228.

\bibitem[Blanchard et~al., 2010]{Blanchard2010Rethinking}
Blanchard, O.~J., Dell'Ariccia, G., and Mauro, P. (2010).
\newblock Rethinking macroeconomic policy.
\newblock {\em Revista de Economia Institucional}, 12(22).

\bibitem[Bulbul, 2013]{bulbul2013determinants}
Bulbul, D. (2013).
\newblock Determinants of trust in banking networks.
\newblock {\em Journal of Economic Behavior \& Organization}, 85:236--248.

\bibitem[Caccioli et~al., 2015]{Caccioli201550}
Caccioli, F., Farmer, J.~D., Foti, N., and Rockmore, D. (2015).
\newblock Overlapping portfolios, contagion, and financial stability.
\newblock {\em Journal of Economic Dynamics and Control}, 51:50 -- 63.

\bibitem[Calorimis and Kahn, 1991]{Calorimis1991}
Calorimis, C.~W. and Kahn, C.~M. (1991).
\newblock The role of demandable debt in structuring optimal banking
  arrangements.
\newblock {\em American Economic Review}, 81(3).

\bibitem[Capponi and Chen, 2015]{Capponi2015152}
Capponi, A. and Chen, P.-C. (2015).
\newblock Systemic risk mitigation in financial networks.
\newblock {\em Journal of Economic Dynamics and Control}, 58:152 -- 166.

\bibitem[Catullo et~al., 2015]{Catullo201578}
Catullo, E., Gallegati, M., and Palestrini, A. (2015).
\newblock Towards a credit network based early warning indicator for crises.
\newblock {\em Journal of Economic Dynamics and Control}, 50:78 -- 97.
\newblock Crises and ComplexityComplexity Research Initiative for Systemic
  InstabilitieS (CRISIS) Workshop 2013.

\bibitem[Chen et~al., 2014]{chen2014money}
Chen, S., Wang, Y., Li, K., and Wu, J. (2014).
\newblock Money creation process in a random redistribution model.
\newblock {\em Physica A: Statistical Mechanics and its Applications},
  394:217--225.

\bibitem[Delli~Gatti and Desiderio, 2015]{Gatti2015}
Delli~Gatti, D. and Desiderio, S. (2015).
\newblock Monetary policy experiments in an agent-based model with financial
  frictions.
\newblock {\em Journal of Economic Interaction and Coordination},
  10(2):265--286.

\bibitem[Delli~Gatti et~al., 2010]{DelliGatti20101627}
Delli~Gatti, D., Gallegati, M., Greenwald, B., Russo, A., and Stiglitz, J.~E.
  (2010).
\newblock The financial accelerator in an evolving credit network.
\newblock {\em Journal of Economic Dynamics and Control}, 34(9):1627 -- 1650.
\newblock Computational perspectives in economics and finance: Methods,dynamic
  analysis and policy modeling.

\bibitem[Diamond and Rajan, 2001]{diamond2001liquidity}
Diamond, D.~W. and Rajan, R.~G. (2001).
\newblock Liquidity risk, liquidity creation, and financial fragility: A theory
  of banking.
\newblock {\em Journal of political Economy}, 109(2):287--327.

\bibitem[Fama, 1980]{fama1980banking}
Fama, E.~F. (1980).
\newblock Banking in the theory of finance.
\newblock {\em Journal of monetary economics}, 6(1):39--57.

\bibitem[Fischer and Riedler, 2014]{Fischer201495}
Fischer, T. and Riedler, J. (2014).
\newblock Prices, debt and market structure in an agent-based model of the
  financial market.
\newblock {\em Journal of Economic Dynamics and Control}, 48:95 -- 120.

\bibitem[Fullwiler, 2008]{fullwiler2008modern}
Fullwiler, S.~T. (2008).
\newblock Modern central bank operations--the general principles.

\bibitem[Gabbi et~al., 2015]{Gabbi2015117}
Gabbi, G., Iori, G., Jafarey, S., and Porter, J. (2015).
\newblock Financial regulations and bank credit to the real economy.
\newblock {\em Journal of Economic Dynamics and Control}, 50:117 -- 143.
\newblock Crises and ComplexityComplexity Research Initiative for Systemic
  InstabilitieS (CRISIS) Workshop 2013.

\bibitem[Galbiati and Soramaoki, 2011]{Galbiati2011859}
Galbiati, M. and Soramaoki, K. (2011).
\newblock An agent-based model of payment systems.
\newblock {\em Journal of Economic Dynamics and Control}, 35(6):859 -- 875.

\bibitem[Gilles et~al., 2015]{Gilles2015375}
Gilles, R.~P., Lazarova, E.~A., and Ruys, P.~H. (2015).
\newblock Stability in a network economy: The role of institutions.
\newblock {\em Journal of Economic Behavior \& Organization}, 119:375 -- 399.

\bibitem[Goodhart et~al., 2016]{goodhart2016macro}
Goodhart, C., Romanidis, N., Tsomocos, D.~P., and Shubik, M. (2016).
\newblock Macro-modelling, default and money.
\newblock {\em Said Business School WP 2016-18}.

\bibitem[Gorton and Pennacchi, 1990]{gorton1990financial}
Gorton, G. and Pennacchi, G. (1990).
\newblock Financial intermediaries and liquidity creation.
\newblock {\em The Journal of Finance}, 45(1):49--71.

\bibitem[Grilli et~al., 2015]{Grilli2015}
Grilli, R., Tedeschi, G., and Gallegati, M. (2015).
\newblock Markets connectivity and financial contagion.
\newblock {\em Journal of Economic Interaction and Coordination},
  10(2):287--304.

\bibitem[Hall, 1983]{hall1983optimal}
Hall, R.~E. (1983).
\newblock Optimal fiduciary monetary systems.
\newblock {\em Journal of Monetary Economics}, 12(1):33--50.

\bibitem[He et~al., 2016]{He20161}
He, J., Sui, X., and Li, S. (2016).
\newblock An endogenous model of the credit network.
\newblock {\em Physica A: Statistical Mechanics and its Applications}, 441:1 --
  14.

\bibitem[Huang et~al., 2010]{Huang20101105}
Huang, W., Zheng, H., and Chia, W.-M. (2010).
\newblock Financial crises and interacting heterogeneous agents.
\newblock {\em Journal of Economic Dynamics and Control}, 34(6):1105 -- 1122.

\bibitem[Iori et~al., 2006]{Iori2006525}
Iori, G., Jafarey, S., and Padilla, F.~G. (2006).
\newblock Systemic risk on the interbank market.
\newblock {\em Journal of Economic Behavior \& Organization}, 61(4):525 -- 542.

\bibitem[Iori et~al., 2015]{Iori201598}
Iori, G., Mantegna, R.~N., Marotta, L., MiccichÃ¨, S., Porter, J., and
  Tumminello, M. (2015).
\newblock Networked relationships in the e-mid interbank market: A trading
  model with memory.
\newblock {\em Journal of Economic Dynamics and Control}, 50:98 -- 116.
\newblock Crises and ComplexityComplexity Research Initiative for Systemic
  InstabilitieS (CRISIS) Workshop 2013.

\bibitem[Jakab and Kumhof, 2015]{jakab2015banks}
Jakab, Z. and Kumhof, M. (2015).
\newblock Banks are not intermediaries of loanable funds--and why this matters.
\newblock {\em Bank of England working paper}.

\bibitem[Klimek et~al., 2015]{Klimek2015144}
Klimek, P., Poledna, S., Farmer, J.~D., and Thurner, S. (2015).
\newblock To bail-out or to bail-in? answers from an agent-based model.
\newblock {\em Journal of Economic Dynamics and Control}, 50:144 -- 154.
\newblock Crises and ComplexityComplexity Research Initiative for Systemic
  InstabilitieS (CRISIS) Workshop 2013.

\bibitem[Krause and Giansante, 2012]{Krause2012583}
Krause, A. and Giansante, S. (2012).
\newblock Interbank lending and the spread of bank failures: A network model of
  systemic risk.
\newblock {\em Journal of Economic Behavior \& Organization}, 83(3):583 -- 608.
\newblock The Great Recession: motivation for re-thinking paradigms in
  macroeconomic modeling.

\bibitem[Ladley, 2013]{Ladley20131384}
Ladley, D. (2013).
\newblock Contagion and risk-sharing on the inter-bank market.
\newblock {\em Journal of Economic Dynamics and Control}, 37(7):1384 -- 1400.

\bibitem[Lux, 2015]{Lux2015A11}
Lux, T. (2015).
\newblock Emergence of a core-periphery structure in a simple dynamic model of
  the interbank market.
\newblock {\em Journal of Economic Dynamics and Control}, 52:A11 -- A23.

\bibitem[Lux, 2016]{Lux201636}
Lux, T. (2016).
\newblock A model of the topology of the bank â€“ firm credit network and
  its role as channel of contagion.
\newblock {\em Journal of Economic Dynamics and Control}, 66:36 -- 53.

\bibitem[MartÃ­nez-Jaramillo et~al., 2010]{Martinez20102358}
MartÃ­nez-Jaramillo, S., PÃ©rez, O.~P., Embriz, F.~A., and Dey, F. L.~G.
  (2010).
\newblock Systemic risk, financial contagion and financial fragility.
\newblock {\em Journal of Economic Dynamics and Control}, 34(11):2358 -- 2374.
\newblock Special Issue: 2008 Annual Risk Management Conference held in
  Singapore during June 30 - July 2, 2008.

\bibitem[Matsuoka, 2012]{Matsuoka20121673}
Matsuoka, T. (2012).
\newblock Imperfect interbank markets and the lender of last resort.
\newblock {\em Journal of Economic Dynamics and Control}, 36(11):1673 -- 1687.

\bibitem[Riccetti et~al., 2013]{Riccetti20131626}
Riccetti, L., Russo, A., and Gallegati, M. (2013).
\newblock Leveraged network-based financial accelerator.
\newblock {\em Journal of Economic Dynamics and Control}, 37(8):1626 -- 1640.
\newblock Rethinking Economic Policies in a Landscape of Heterogeneous Agents.

\bibitem[Ricks, 2016]{ricks2016money}
Ricks, M. (2016).
\newblock {\em The Money Problem: rethinking financial regulation}.
\newblock University of Chicago Press.

\bibitem[Romer, 2016]{romer2016trouble}
Romer, P. (2016).
\newblock The trouble with macroeconomics.
\newblock {\em September, forthcoming in The American Economist}.

\bibitem[Rule, 2015]{rule2015understanding}
Rule, G. (2015).
\newblock Understanding the central bank balance sheet.
\newblock {\em Handbooks}.

\bibitem[Schumpeter, 1954]{Schumpeter1954}
Schumpeter, J.~A. (1954).
\newblock {\em History of Economic Analysis}.
\newblock New York: Oxford University Press.

\bibitem[Shubik, 2011]{shubik2011note}
Shubik, M. (2011).
\newblock A note on accounting and economic theory: Past, present, and future.
\newblock {\em Accounting, Economics, and Law}, 1(1).

\bibitem[Shubik and Smith, 2016]{shubik2016guidance}
Shubik, M. and Smith, E. (2016).
\newblock {\em The Guidance of an Enterprise Economy}.
\newblock Mit Press.

\bibitem[Stiglitz and Gallegati, 2011]{stiglitz2011heterogeneous}
Stiglitz, J.~E. and Gallegati, M. (2011).
\newblock Heterogeneous interacting agent models for understanding monetary
  economies.
\newblock {\em Eastern Economic Journal}, 37(1):6--12.

\bibitem[Temizsoy et~al., 2015]{Temizsoy2015118}
Temizsoy, A., Iori, G., and Montes-Rojas, G. (2015).
\newblock The role of bank relationships in the interbank market.
\newblock {\em Journal of Economic Dynamics and Control}, 59:118 -- 141.

\bibitem[Teteryatnikova, 2014]{Teteryatnikova2014186}
Teteryatnikova, M. (2014).
\newblock Systemic risk in banking networks: Advantages of
  â€œtieredâ€� banking systems.
\newblock {\em Journal of Economic Dynamics and Control}, 47:186 -- 210.

\bibitem[Tobin, 1963]{tobin1963commercial}
Tobin, J. (1963).
\newblock Commercial banks as creators of ‘money”, cowles foundation
  discussion papers no. 159.
\newblock {\em Cowles Foundation for Research in Economics, New Haven,
  Connecticut, Yale University}.

\bibitem[Valencia, 2014]{Valencia201420}
Valencia, F. (2014).
\newblock Monetary policy, bank leverage, and financial stability.
\newblock {\em Journal of Economic Dynamics and Control}, 47:20 -- 38.

\bibitem[Werner, 2014a]{werner2014can}
Werner, R.~A. (2014a).
\newblock Can banks individually create money out of nothing?â€”the
  theories and the empirical evidence.
\newblock {\em International Review of Financial Analysis}, 36:1--19.

\bibitem[Werner, 2014b]{werner2014banks}
Werner, R.~A. (2014b).
\newblock How do banks create money, and why can other firms not do the same?
  an explanation for the coexistence of lending and deposit-taking.
\newblock {\em International Review of Financial Analysis}, 36:71--77.

\bibitem[Xiong et~al., 2017]{Xiong2017425}
Xiong, W., Fu, H., and Wang, Y. (2017).
\newblock Money creation and circulation in a credit economy.
\newblock {\em Physica A: Statistical Mechanics and its Applications}, 465:425
  -- 437.

\bibitem[Xu et~al., 2016]{Xu2016131}
Xu, T., He, J., and Li, S. (2016).
\newblock A dynamic network model for interbank market.
\newblock {\em Physica A: Statistical Mechanics and its Applications}, 463:131
  -- 138.

\end{thebibliography}

\end{document}